\documentclass[useAMS,usegraphicx,usenatbib]{mn2e}
\usepackage{amssymb}
\title[A \emph{Spitzer}/IRAC Characterization of Galactic AGB and RSG Stars]{A \emph{Spitzer}/IRAC Characterization of Galactic AGB and RSG Stars}

\author[Reiter et al.]{Megan Reiter$^{1}$, Massimo Marengo$^{2}$,
Joseph L. Hora$^{3}$, and Giovanni G. Fazio$^{3}$ \\
$^{1}$ Steward Observatory, University of Arizona, Tucson,
  AZ 85721, USA \\
$^{2}$ Department of Physics and Astronomy, Iowa State
  University, Ames, IA 50011, USA \\
$^{3}$ Harvard-Smithsonian Center for Astrophysics, 60 Garden
  St. Cambridge, MA 02138-1516, USA}

\begin{document}

\date{Accepted 2014 Jul ??. Received 2014 Jul ??; in original form 2014 Jul ??}

\pagerange{\pageref{firstpage}--\pageref{lastpage}} \pubyear{2014}

\maketitle

\label{firstpage}

%%=============================================================================

\begin{abstract}
We present new \emph{Spitzer}/IRAC observations of 55 dusty Long 
Period Variables (LPVs, 48 AGB and 6 RSG stars) in the Galaxy that 
have different chemistry, variability type, and mass-loss rate. 
O-rich AGB stars (including intrinsic S-type) tend to have redder 
[3.6]$-$[8.0] colors than carbon stars for a given [3.6]$-$[4.5] color 
due to silicate features increasing the flux in the 8.0 \micron\ IRAC band. 
For colors including the 5.8 \micron\ band, carbon stars separate into two 
distinct sequences, likely due to a variable photospheric C$_3$ feature that 
is only visible in relatively unobscured, low mass-loss rate sources. 
Semiregular variables tend to have smaller IR excess in [3.6]$-$[8.0] 
color than Miras, consistent with the hypothesis that semiregular variables 
lose mass discontinuously. Miras have redder colors for longer periods while 
semiregular variables do not. Galactic AGB stars follow the period-luminosity 
sequences found for the Magellanic Clouds. Mira variables fall along the 
fundamental pulsation sequence, while semiregular variables are mostly on overtone sequences. We also derive a relationship between mass-loss rate and 
[3.6]$-$[8.0] color. The fits are similar in shape to those found by other 
authors for AGBs in the LMC, but discrepant in overall normalization, likely 
due to different assumptions in the models used to derive mass-loss rates. 
We find that IR colors are not unique discriminators of chemical type, 
suggesting caution when using color selection techniques to infer the 
chemical composition of AGB dust returned to the ISM.
\end{abstract}


%%=============================================================================

\section{Introduction}\label{s:intro}

The Asymptotic Giant Branch (AGB) is the last evolutionary phase of low and intermediate mass stars ($M < 8$ M$_{\odot}$), before the brief post-AGB phase that leads them to become white dwarfs. 
AGB stars are characterised by complex nucleosynthesis, high mass-loss rates, and variability. They are the source of a large fraction of the mass returned by stars to the Interstellar Medium \citep[ISM,][]{sed94}. In particular, they are believed to be the primary `dust factories' in galaxies \citep{gehrz1989,boyer2012}, even though recent observations have re-evaluated the role played by supernovae \citep[see e.g.][]{matsuura2011,dwek2011}. It is clear, however, that understanding mass-loss in AGB stars is instrumental for modeling the chemical evolution of galaxies. 

The composition of the material returned from AGB stars to the ISM depends on the chemistry of the star, which is determined by its initial mass and evolutionary history. The latter is crucially regulated by mass loss. 
Most stars enter the AGB with an intrinsic C/O abundance ratio less than one (O-rich). In stars of $\sim 1.5 - 4$ M$_{\odot}$ \citep{str95,str97} however, several dredge-up events following thermal pulses may drive the C/O ratio above unity. These stars become the so-called carbon stars. 
In stars above $\sim 4$ M$_{\odot}$, however, carbon is destroyed by nuclear burning at the base of the convective envelope \citep[Hot Bottom Burning;][]{smi85,boo92} preventing it from reaching the surface of the star where it could be observed. These stars never transition to the carbon star phase. 
Mass limits constraining the formation of carbon stars depend on metallicity; low metallicity environments (e.g. the Magellanic Clouds) form carbon stars at lower masses. 
% (see discussion in Section~4).
Intrinsic S-type stars share many characteristics with carbon stars such as similar masses and a rich s-element chemistry from many dredge-up events, but they lack sufficient carbon to drive the C/O ratio above one. 

\begin{figure*}
\centering
\includegraphics[angle=0,scale=0.325]{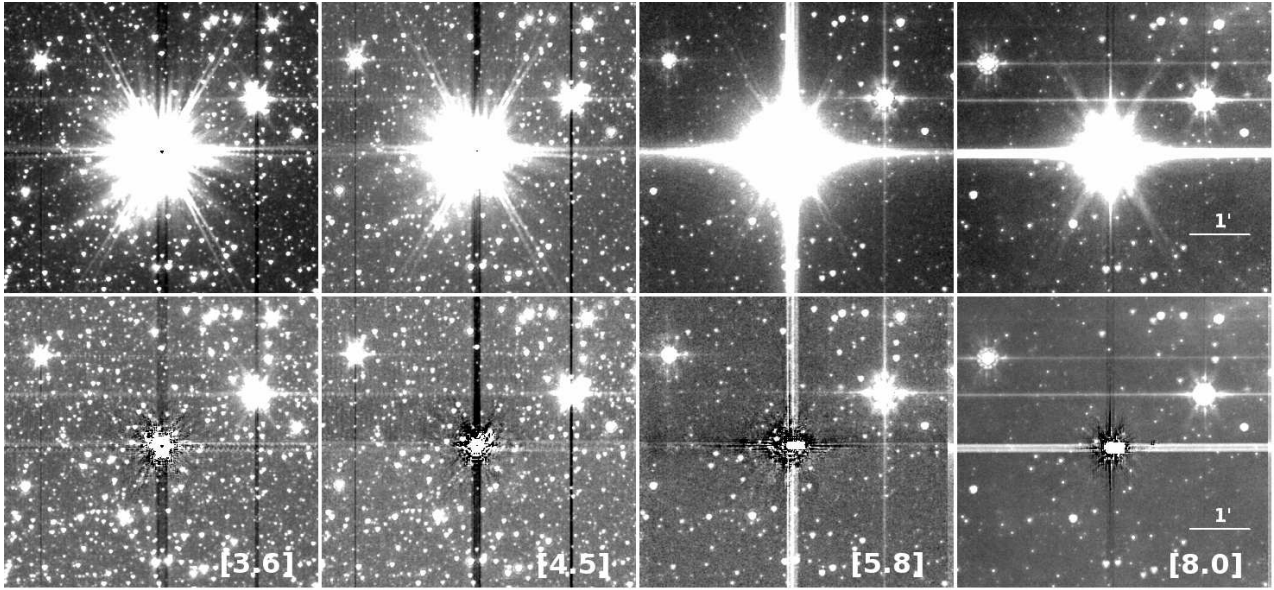}
\caption{Images of O-rich AGB star CZ Ser in each of the IRAC bands
  before PSF subtraction (top panels) and after PSF subtraction
  (bottom panels). The horizontal and vertical bands are un-subtracted
  electronic artefacts that are ignored in the fit. The PSF was fit to 
  unsaturated diffraction spikes and rings, outside the saturation radius 
  ($\sim 20$\arcsec\ for CZ Ser). The field of view
  of each panel is $\sim 5'$.}\label{fig:photometry_pics}
\end{figure*}

Stars with masses $\sim 8-25$ M$_{\odot}$ do not enter the AGB phase, but become Red Supergiants (RSGs) on the way to ending their lives as Type~II supernovae \citep[e.g.][]{mas03,van03}. RSGs do not experience third dredge up episodes and as such they are characterised by an O-rich chemistry \citep[e.g.][]{ver09}. More luminous than AGB stars, but with similar spectral types, they are also subject to variability and mass loss. Their dust production rates are comparable to those of AGB stars but their overall yield to the ISM is lower, due to RSGs being more rare than their lower mass counterparts in the AGB \citep[e.g.][]{boyer2012}.

AGB and RSG stars lie close to an instability region of the Hertzsprung-Russell diagram and as such they exhibit long period variability of Mira, semiregular, and irregular type with periods on the order of hundreds to thousands of days. 
Period-luminosity diagrams of long period variables (LPVs) in the Magellanic clouds (LMC and SMC) reveal sequences that appear to correspond to the variable pulsation mode, binarity and other as yet unknown characteristics \citep{w10}. 
These diagrams can be a powerful tool for studying the pulsation mode for a large population of variable stars. 
This is especially true when the available light curves are sparse and poorly sampled, making it difficult to detect additional periodicity (e.g. overtone pulsations) in the amplitude modulations of the light curve. 
The construction of similar diagrams for Galactic AGB stars would be extremely useful, as they would illuminate the role of metallicity in determining the pulsation properties of these variables. Efforts in this sense depend critically on accurate and precise distance measurements for Galactic LPVs \citep{whitelock2008}. 

Pulsations and mass loss are inextricably linked in AGB and RSG stars \citep[see e.g.][]{willson2000} since radial pulsations aide in the formation of the dust-driven winds responsible for the characteristic high mass-loss rates of AGB stars ($10^{-8}$ M$_{\odot}$ yr$^{-1}$ up to $10^{-4}$ M$_{\odot}$ yr$^{-1}$, e.g. \citealt{woo83,woo92,van99}, as compared with $10^{-9}$ M$_{\odot}$ yr$^{-1}$ to $10^{-7}$ M$_{\odot}$ yr$^{-1}$, e.g. \citealt{mau06,dup09} for Red Giant Branch stars). 
The dusty wind that results from these mass-loss processes, before elements synthesized in AGB stars are released into the ISM, leads to the formation of dusty cocoons enshrouding the star. 
These circumstellar envelopes obscure optical radiation, making observation of mass-losing AGB stars difficult at visible wavelengths. However, thermal radiation from the dust makes AGB stars extremely luminous in the infrared (IR). 
The InfraRed Array Camera \citep[IRAC,][]{faz04} on board the \emph{Spitzer Space Telescope} \citep{wer04} operates in the mid-IR and is an ideal instrument for characterizing AGB stars according to their chemical composition, mass-loss rate, and variability class. 
IRAC's four channel imaging during the \emph{Spitzer} cryogenic mission was especially well-suited to observe the prominent dust and molecular spectral features in AGB stars -- for example, H$_2$O, SiO, CO$_2$, and CO among others for O-rich stars and C$_2$H$_2$, HCN, CS, and C$_3$ for carbon stars. 

Due to their intrinsic brightness and red colors, AGB stars are easily detected by IRAC and have been found in numerous IRAC surveys of the Milky Way (e.g. GLIMPSE, \citealt{ben03,chu09}) and Local Group galaxies (e.g. SAGE, \citealt{mei06,blum2006,boyer2011}). 
However, there have been no studies specifically designed to characterise in the IRAC bands a sample of Galactic AGB stars with well determined chemical type, variability, and mass-loss rate. 
To fill this gap, we designed a program to observe a nearby sample of Galactic AGB stars with IRAC in order to better identify AGB stars in IRAC surveys, to facilitate Galactic population synthesis studies, and to better understand the chemical evolution of the diffuse matter in the ISM. 

We have observed 48 Galactic AGB stars, representing each of the main types of AGB stars -- O-rich, S-type, carbon stars, Mira, semiregular, and irregular variables as well as 6 RSGs of semiregular and irregular variability type. 
We outline our target selection in Section 2. Section 3 describes the data reduction pipeline and how the photometry was derived.  
We present the IRAC colors of AGB stars in Section 4 and discuss period, magnitude and color relations in Section 5. 
Relationships with the mass-loss rate are explored in Section 6. 
We summarise our conclusions in Section 7. 

%%=============================================================================

\section{Target Selection}\label{s:targets}

Our target list was selected from a number of Galactic AGB star catalogs
available in the literature \citep[and references therein]{lou93,ker96,ade98,her05,gua06} with the intention of
representing all main types of AGB stars. Two main constraints limited
our choice of targets: (1) the availability of distance
estimates and (2) a $K$ band magnitude $\ga 0$, to prevent excessive
saturation with IRAC. While our target list is not a statistically significant
sample of the Galactic AGB population, the sample is sufficiently large 
and diverse to allow a general study of the spectral energy 
distribution (SED) of Galactic AGB stars in the \emph{Spitzer}/IRAC bands, 
and its overall dependence on their chemical, physical, and variability
characteristics.

Distances for our target stars were derived either from interferometric 
observations \citep[e.g.][]{van02,zha12}, models of
radio emission and/or bolometric luminosity \citep{lou93,olo02,gua06,gua08,sch13}, and, in the absence of either of those, astrometric methods (\citealt{van07}, adopting the latest corrections of Hipparcos distances). 
The distance of our selected AGB targets varies from 0.14 to 1.85~kpc (with one
single carbon star that has a distance of 4.21~kpc).
The distances to previously observed supergiants have been derived with 
different methods found in the literature (see notes on Table~\ref{orig_stars}). 

The complete sample is listed in Table~\ref{orig_stars}, and is
divided in three main categories: 22 O-rich AGB stars (M~III spectral
type), 7 intrinsic S stars, 19 carbon stars, and 6 supergiants. 
In each category we have
a similar number of Mira, semiregular and irregular variables, with
periods ranging from 50 to 822 days. 
Information about the period and variability type of the targets was
obtained from the General Catalog of Variable Stars \citep[GCVS,][]{sam12}. 
Estimated mass-loss rates were obtained from radio
observations of CO or HCN in the outer envelope, or fitting detailed 
radiative transfer models of the circumstellar emission to the infrared SED \citep[see][and references therein]{lou93,olo02,gua06}. 
The mass-loss rates of our target stars range from
$10^{-8}$ to $10^{-4}$~$M_{\sun}$~yr$^{-1}$ in each category.
Mass-loss rates for the supergiants 
were estimated by various methods (see notes in Table~\ref{orig_stars}). 
Near-IR photometry was obtained (for
random epochs) from the 2MASS catalog \citep{skr06}.
While higher quality near-IR photometry exists for some of the stars in our sample, we chose to adopt 2MASS photometry for uniformity across the sample. The uncertainty of the 2MASS magnitudes are already smaller than the amplitude of the infrared variability of our LPVs. 
In absence of complete light-curves for all stars in the sample from which to calculate average magnitudes, this obviates the need for single epoch photometry with better precision than 2MASS.

Given the variability of the sources, we requested two epochs
(six months apart, as constrained by the \textit{Spitzer} visibility
windows) for each target, in order to check for variations in the IRAC
photometry on time-scales of several months.

%%=============================================================================

\section{Observations and Data Reduction}\label{s:obs}

We restricted our target list to nearby AGB stars with estimates of 
their distance. 
Given the intrinsic luminosity ($\sim 10^4 \ L_\odot$) and
red colors of AGB stars, these targets saturate the
IRAC detectors even at the shortest IRAC subarray frame times. While
saturated images are not suitable for aperture photometry, reliable
photometry can be recovered with Point Spread Function (PSF)
fitting, as long as at least part of the PSF is not saturated. To
compromise between the need to limit the amount of saturation in
our images and, at the same time, fill enough of the IRAC field-of-view 
with high S/N unsaturated portions of the PSF, we adopted the 2~sec
full-frame IRAC Astronomical Observation Template. To allow for
efficient removal of outliers (bad pixels and cosmic rays), and
sufficient spatial sampling of the unsaturated PSF optical features
(diffraction spikes and rings), we observed each target using the 
5-point small scale Gaussian dither pattern.

The data were acquired between 2006 June 2 and 2008 June 20 as part of
the IRAC Guaranteed Time, with PID 30411. Starting with
the Basic Calibrated Data (BCD) produced by the \textit{Spitzer}
Science Center (SSC) pipeline version S14.4.0, we generated a
mosaiced image for each source, using our own post-BCD software
IRACproc \citep{sch06}.  IRACproc is an add-on to SSC's MOPEX
mosaicing software, that applies more sophisticated outlier rejection
criteria.  With the optimised
parameters from IRACproc, MOPEX defines a fiducial image frame that
contains the coordinates for the constituent frames, removes transients
using temporal outlier rejection and then interpolates the frames to
produce a mosaic image. The final mosaic is scaled to 0.863$''$/pixel
which is $1/\sqrt{2}$ of the IRAC pixel scale (half of the IRAC native 
pixel area), to provide ideal (Nyquist) sampling even at the
shortest IRAC wavelengths.

\begin{figure*}
\centering
\includegraphics[angle=90,scale=0.60]{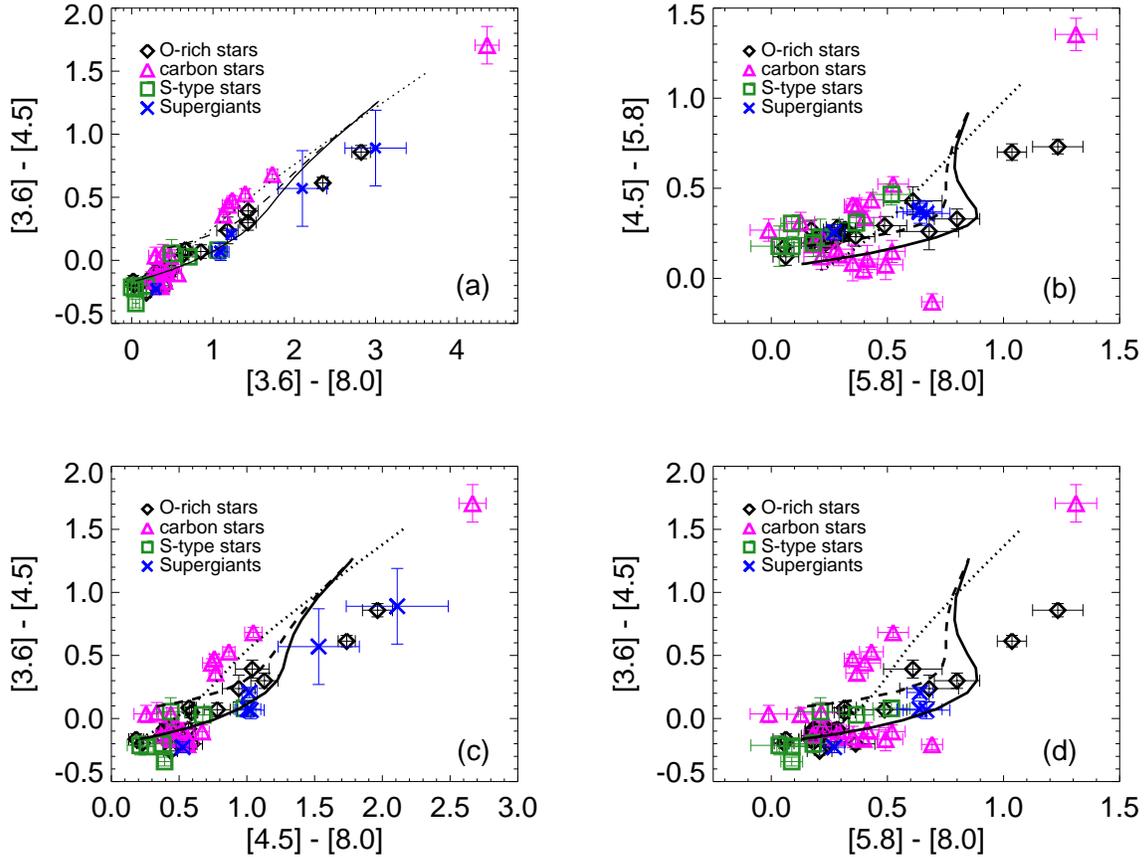}
\caption{Color-Color diagrams (average of the two epochs) of the observed AGB
  stars. O-rich stars are marked with diamonds, carbon stars with
  triangles, intrinsic S-type stars with squares and supergiants with
  crosses. Model tracks from \citet{gro06} for three spectral types are 
  overplotted. Carbon star models are plotted with dotted lines, and models 
  with an M0~III and M10~III central star and silicate dust are plotted with 
  solid and dashed lines, respectively.}\label{fig:all_color}
\end{figure*}

Table~\ref{phot_table} lists the measured IRAC Vega magnitudes for all
sources. The adopted PSF fitting technique, developed specifically for
heavily saturated IRAC images, is described in \citet{mar09}. As an
example of this procedure, Figure~\ref{fig:photometry_pics} shows the
O-rich AGB star CZ~Ser, shown before and after the PSF is fit and
subtracted from the target. We used a high dynamic range image of the
IRAC PSF\footnote{available at the IRSA website: 
http://irsa.ipac.caltech.edu/\\data/SPITZER/docs/irac/calibrationfiles/psfprf/} 
in each band, made from a combination of individual images of
a set of stars with different brightness (Sirius, Vega, Fomalhaut,
$\epsilon$~Eridani, $\epsilon$~Indi, and the IRAC calibrator
BD$+$68 1022).  The PSF intensity is scaled to match the actual IRAC 
observations of Vega in each band, providing an absolute
photometric reference, and is super-sampled over a grid with
0.24$''$/pixel. By fitting the unsaturated parts (diffraction spikes
and PSF ring and ``tails'') of each target star image, we have
measured the flux ratio between the stars in our sample and Vega. 
%, which we have converted to Vega magnitudes for each source. 
The typical uncertainty of this procedure is 2$-$5\% (constrained by
comparing the amplitude of the PSF subtraction residuals with
the background and source photon noise), and is comparable to the
typical absolute aperture photometry of unsaturated stars with IRAC.
The photometry of the six supergiants from the literature 
obtained from \citet{sch07} use the same PSF-fitting technique. 

%%=============================================================================

\begin{figure}
\includegraphics[trim=0mm 0mm 0mm 17.5mm,angle=90,scale=0.35]{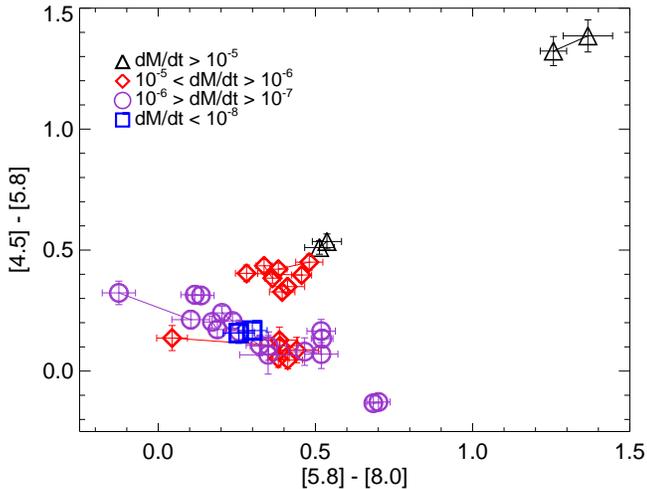}
\caption{Color-color diagram of our carbon stars with different
  symbols for increasing mass-loss rate. For each star we show the
  color for both epochs, connected by a line. Stars with the highest
  mass-loss rates are on the upper branch, with redder [4.5]$-$[5.8]
  color than the lower branch. Stars with lower mass-loss rates are on
  the lower track, indicating depressed flux at
  5.8~\micron.}\label{fig:carbon_color}
\end{figure}

\section{IRAC Colors of Galactic AGB Stars}\label{s:colors}

Ground-based mid-IR colors have been used extensively to characterise
the physical and chemical properties of the circumstellar envelopes of
evolved stars (see e.g. \citealt{mar99} and references therein). The
IRAC bands offer a similar opportunity, as they span a wavelength
range rich in strong molecular and dust features, and they are free
from variable telluric absorption lines. These features correlate with
the chemical signature of the circumstellar environment, dust
mineralogy, and abundance in the stellar wind. 
Note that in the IR, interstellar extinction is dramatically reduced with respect to the visible; as a consequence the mid-IR and near-IR colors of our sources are mainly determined by photospheric and circumstellar features. 

Figure~\ref{fig:all_color} shows a number of IRAC color-color diagrams
for all target sources. Each source is plotted once, using the average
color of the two observed epochs. The sources tend to be organized on
a sequence of increasing excess colors, matching the overall
distribution of colors found by \citet{mar07} using synthetic IRAC
photometry derived from Infrared Space Observatory Short Wavelength
Spectrometer \citep[ISO SWS,][]{val96} spectra (from
\citealt{slo03}). The largest dispersion is in the [3.6]$-$[8.0] color
(panel \emph{a}). Radiative transfer models (see e.g. \citealt{gro06})
show that this sequence is related to increasing amounts of
circumstellar dust, indicating larger mass-loss rates (assuming wind
velocities, dust opacities, and a uniform dust-to-gas mass ratio 
across the sample). We explore quantitatively the correlation between
mass-loss rates and IRAC colors in section~\ref{s:comparison}.

Stars extending along the infrared excess sequence tend to separate
according to their circumstellar chemistry. O-rich and intrinsic
S-type AGB stars, as well as the supergiants, tend to have a redder
[3.6]$-$[8.0] color (for the same [3.6]$-$[4.5] color), due to the
presence of silicate features dramatically increasing their flux in
the 8.0~\micron\ band. This separation is only effective for
sources with moderate infrared excess, e.g. $[3.6] - [8.0] \ga 1$ mag. 

The \citet{gro06} models we selected miss the location of the most 
extreme O-rich AGB stars in all of the color-color diagrams. 
This may be due to the choice of dust chemistry in the model. 
In Figure~\ref{fig:all_color}, we plot models using a mixed chemistry 
-- 60\% aluminum oxide (amorphous porous  Al$_2$O$_3$) and 40\% silicates \citep[see][for details]{gro06}. 
More recent papers have shown that the choice of optical constants affects the predicted flux of the silicate feature near 10 \micron\ \citep[see][]{gro09}. 
This suggests that the particular silicate opacities employed in the model may lead to the inaccurate determination of the dust mass-loss rate. 

Plots including colors with the 5.8~\micron\ band (panels \emph{b} and
\emph{d}) instead show an unexpected separation of carbon stars into
two distinct branches. Both branches show similar [5.8]$-$[8.0]
excess, but the upper branch continues to get progressively redder in
both [4.5]$-$[5.8] and [3.6]$-$[4.5] colors, whereas stars in the
lower branch tend to get bluer in the [4.5]$-$[5.8] color. This
divergence suggests the presence of a broad absorption feature at
5.8~\micron.  Looking at just the carbon stars in [5.8]$-$[8.0] versus
[4.5]$-$[5.8] color and using different symbols to indicate increasing
mass-loss rate (Figure~\ref{fig:carbon_color}), we see that the
separation is related to the mass-loss rate listed in
Table~\ref{orig_stars}. Stars with higher mass-loss rates are on the upper,
red branch, while stars with lower mass-loss rates populate the lower,
blue track. This dichotomy suggests that the feature responsible for
the 5.8~\micron\ absorption is likely photospheric, being masked by
thicker dusty envelopes in sources with higher mass-loss rates.

\begin{figure}
$\begin{array}{c}
\includegraphics[trim=0mm 10mm 0mm 10mm,angle=90,scale=0.30]{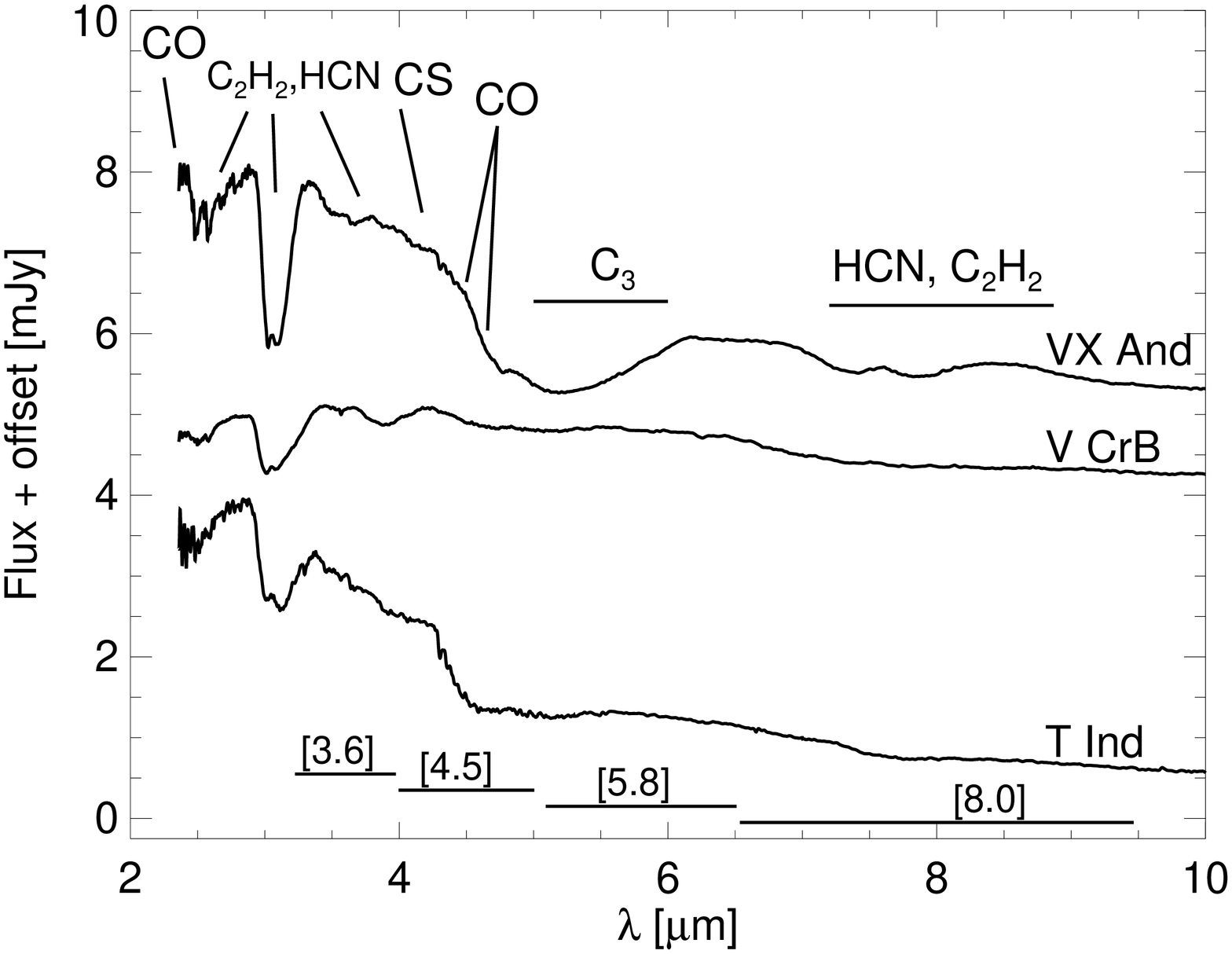} \\
\includegraphics[trim=0mm 10mm 0mm 10mm,angle=90,scale=0.30]{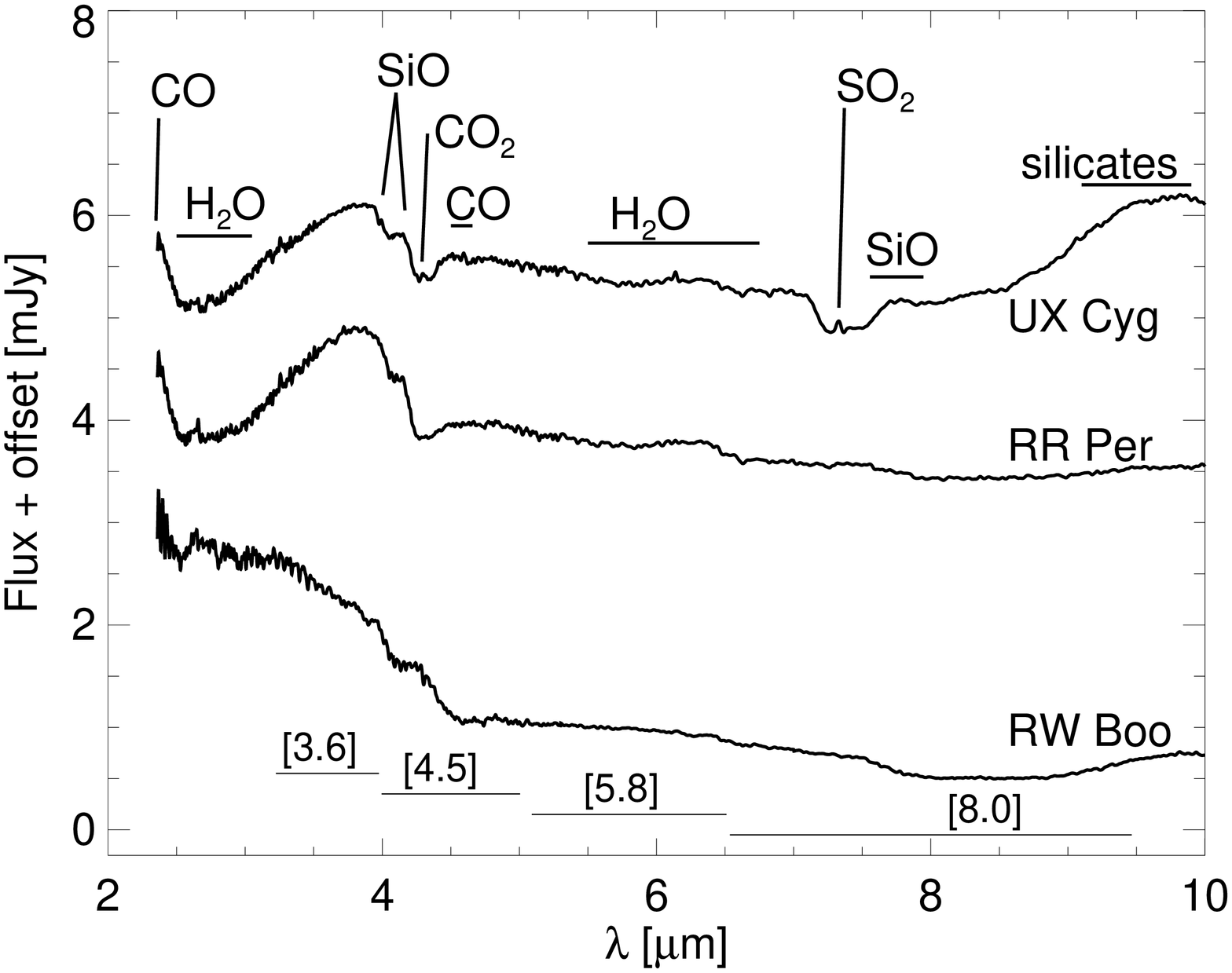} \\
\includegraphics[trim=0mm 10mm 0mm 10mm,angle=90,scale=0.30]{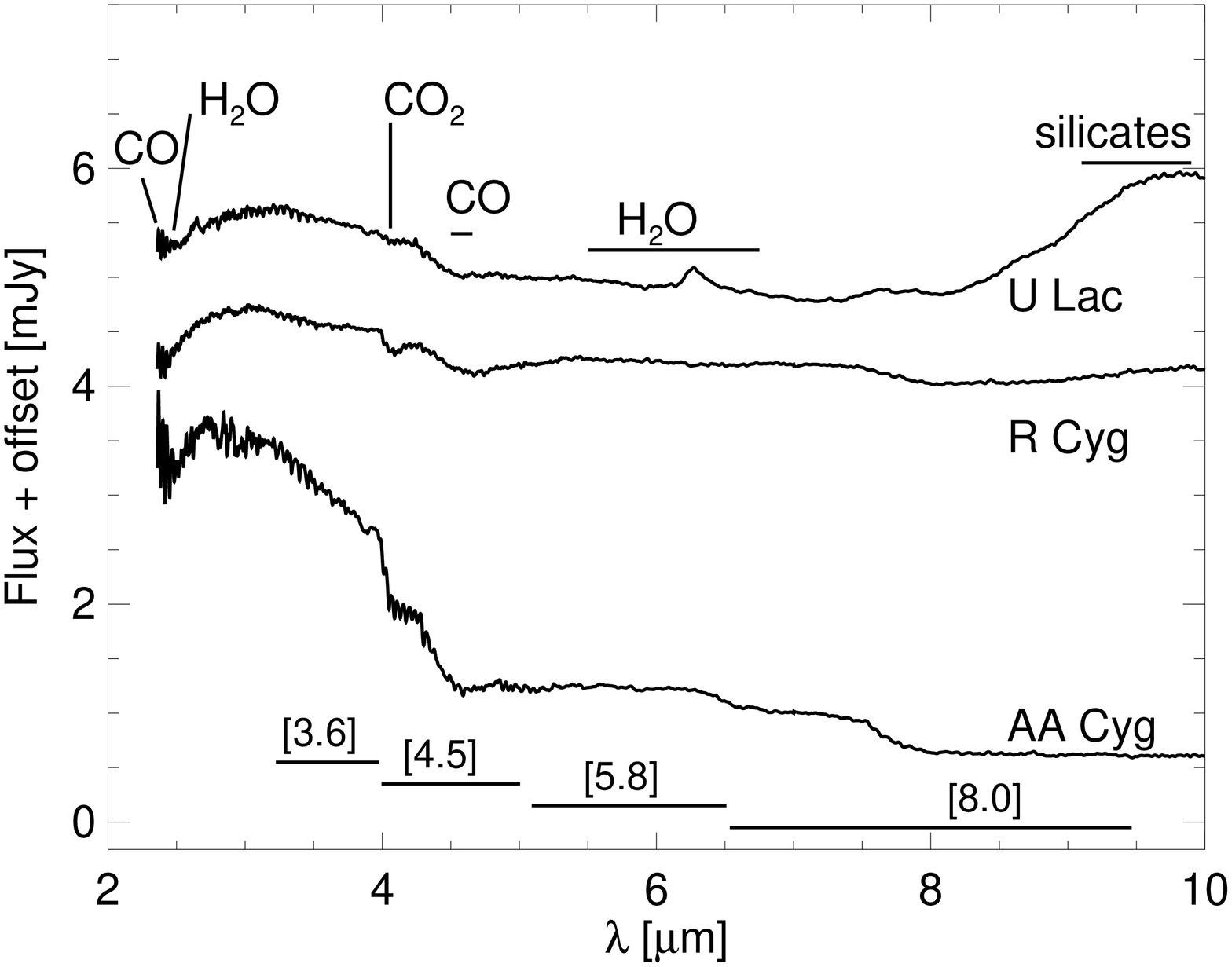} \\
\end{array}$
\caption{ISO SWS spectra of representative stars in our
  sample. \emph{Top:} -- carbon stars; 
  \emph{Center:} -- O-rich AGB stars; 
  \emph{Bottom:} -- two intrinsic S-stars and 
  a red supergiant (AA~Cyg). The IRAC passbands are plotted below the
  spectra and the main photospheric and circumstellar features are
  indicated \citep[following, e.g.][]{aok99,nor04,zha08}.}\label{fig:all_isosws}
\end{figure}

To identify this feature, and to match the observed trends in the IRAC
colors with specific spectral features, we retrieved archival spectra
\citep{slo03} for all of our sources that were observed with ISO
SWS. We have also compared our colors with the synthetic photometry of 
122 AGB and RSG stars of different chemical type, derived by \citet{mar07} 
by convolving available ISO/SWS spectra with the IRAC passbands. 
Some of these sources have been observed by ISO at multiple epochs 
allowing a direct test of the effect of variable spectral 
features on the IRAC colors. Representative examples are shown in
Figure~\ref{fig:all_isosws}. The spectrum of carbon stars along the
blue branch (e.g. VX~And in the top panel) exhibits a strong C$_3$
absorption feature centred at 5~\micron. This feature is not present
in the spectra of stars in the redder branch (e.g. V~CrB and
T~Ind). Based on \citet{slo98}, C$_3$ is assumed to be developing in
the atmospheres of dust poor carbon stars. This feature disappears in
stars with thicker circumstellar envelopes, either because it is
filled by the continuum dust emission, or because the molecule making
the feature is depleted. As noted in \citet{mar07}, this feature
appears to be transient, as some sources observed with ISO SWS in
multiple epochs do not show it in all spectra. This suggests that the
feature variability is related to changes in the C$_3$ abundances in
the stellar atmosphere as the star pulsates, rather than changes in
the dust content of the circumstellar envelope \citep[which is unlikely to
show large scale variations on the short time-scales of the ISO
repeated observations, although such variations are found on longer time-scales, see e.g.][]{whitelock2006}.

The spectra of carbon stars present several other molecular features
in the IRAC passbands (CO, C$_2$H$_2$, HCN and CS). 
C$_2$H$_2$ falls mostly outside the 3.6~\micron\ band, and does not contribute 
significantly to the [3.6]$-$[4.5] color of the carbon stars. 
CO absorption in the 4.5~\micron\ band more likely contributes to the 
negative (as low as $\simeq -0.2$ mag) [3.6]$-$[4.5] color of the carbon 
stars with low overall infrared excess. 
All these molecular absorption features are
filled by dust continuum emission for the redder sources, explaining
the general trend of increasing [3.6]$-$[8.0] color for stars with
larger circumstellar dust content.

The mid-IR spectra of the O-rich sources (middle panel), intrinsic 
S-stars and a red supergiant (bottom panel) are also rich in several molecular
features (including CO, H$_2$O, SiO, CO$_2$ and SO$_2$). Of these,
the features that have the largest effect on the IRAC colors are SiO, 
CO$_2$, and CO bands that can severely depress
the 4.5~\micron\ flux leading to blue [3.6]$-$[4.5] colors (as low as
$\sim -0.4$ mag) for the sources with less overall infrared
excess. Sources with high dust content show the prominent 10~\micron\
silicate feature in emission. As mentioned before, this feature is
partially captured by the 8.0~\micron\ band, causing the [3.6]$-$[8.0] 
color to be redder for higher mass-loss rates. There are no
significant differences between the colors of the individual
types of O-rich sources, as evidenced by the fact that M-type AGB
stars, intrinsic S-stars and red supergiants all trace the same color
sequence.

\begin{figure}
\includegraphics[trim=0mm 0mm 0mm 17.5mm,angle=90,scale=0.35]{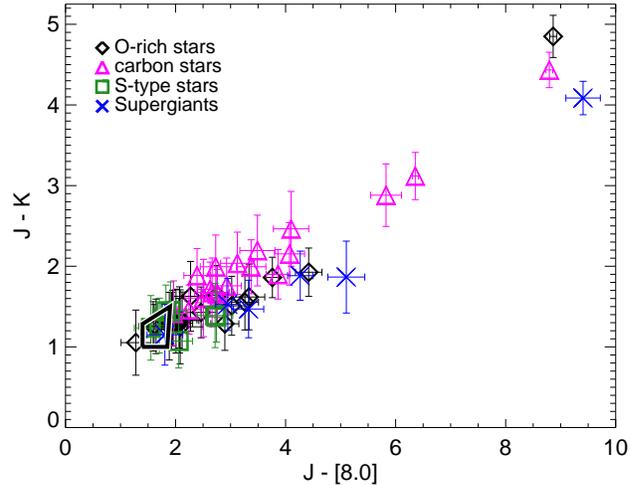}
\caption{Color-magnitude diagram using near-IR photometry from 2MASS
  and Spitzer $8$ \micron. The trapezial box frames anomalous 
  O-rich AGB stars as identified by \citet{boyer2011}. }\label{fig:j-8.0} 
\end{figure}

A number of classification schemes for the chemistry of AGB stars
rely on the combination of IRAC with near-IR colors. Figures~7 and 8
in \citet{boyer2011}, for example, show how O- and C-rich sources in
the SMC and LMC segregate according to their J$-$[8.0]
color. With this scheme, O-rich AGB stars have $J-[8.0] \la 1.4$ mag and C-rich
AGB stars have $J-[8.0] \ga 1.9$ mag, with the so-called ``anomalous'' O-rich
population having intermediate color. We do not see this segregation
for Galactic AGB stars, as shown in our Figure~\ref{fig:j-8.0}, where
C-rich and O-rich AGB stars overlap above $J-[8.0] > 2$ mag and
intrinsic S-type stars (proposed as one possible explanation
for the LMC/SMC anomalous O-rich AGB class) overlap with the range 
of the anomalous O-rich AGB stars (see box in Figure~\ref{fig:j-8.0}).

The fact that we have several O-rich AGB stars that have significantly
redder colors than most O-rich AGB stars in the Magellanic Clouds 
may be the consequence of the different evolutionary paths of the two
AGB populations in galaxies with different metallicity. 
AGB stars are observed to have redder colours at higher metallicities because 
the gas to dust ratio increases \citep{van00}. 
We note that the few O-rich AGB stars in the Magellanic Clouds that have redder 
colours than the Galactic AGB stars in our sample are 
OH/IR stars \citep{woo92,mar04}. 
As noticed by
many authors \citep[see, e.g.][]{fer06,ven12} a low metallicity 
environment favors the creation
of carbon stars early in the evolution of an AGB star, when the mass-loss 
rate (and hence infrared excess) tends to be smaller. According
to this hypothesis, LMC/SMC O-rich AGB stars are restricted to low J$-$[8.0]
excess, with only C-rich stars having redder J$-$[8.0] colors due 
to their greater mass loss. This restriction would not be present in
Galactic O-rich AGB stars, requiring a higher number of thermal pulses to
transition to C-rich chemistry.  Alternatively (though not in our sample), 
the neat color separation observed in the LMC/SMC C- and O-rich AGB stars 
could be explained by the
selection method used to identify the chemical type of the
source. LMC/SMC AGB catalogs are typically based on near-IR colors 
(see e.g. Figure~5 in \citealt{boyer2011}) while the chemistry of our 
targets has been identified spectroscopically. This second possibility 
is supported by the
large range in infrared colors exhibited by spectroscopically
identified C- and O-rich LMC AGB stars (see e.g. \citealt{matsuura2009}). 
Finally, this may be a selection bias in our sample as we have 
selected sources with measurable mass-loss rates, and hence a larger 
infrared excess. 

\begin{figure}
\includegraphics[trim=0mm 0mm 0mm 17.5mm,angle=90,scale=0.35]{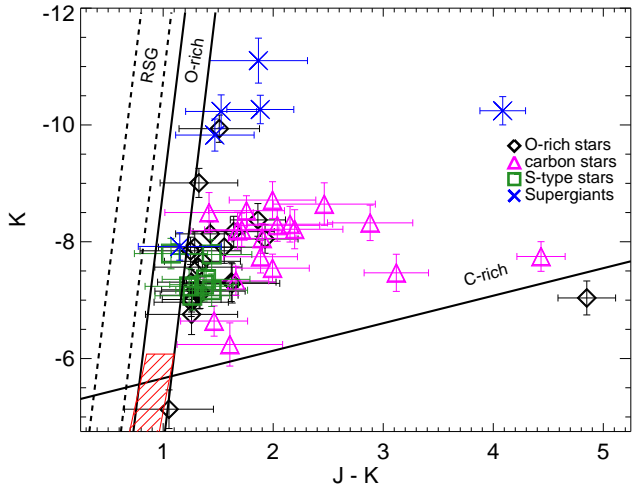}
\caption{Color-magnitude diagram using near-IR photometry from
  2MASS. Lines correspond to the near-IR color cuts used by \citet{boyer2011}.
  The red striped box denotes the RGB. 
}\label{fig:j-k} 
\end{figure}

The two reddest O-rich AGB stars in our sample are KU~And and V1300~Aql. 
Both have red [3.6]$-$[8.0] color and V1300~Aql is the reddest in the 
near-IR (consistent with its fainter J-band magnitude indicative of 
higher circumstellar extinction, 
%$J\approx7$ mag, compared to KU~And, $J\approx 3$ mag, 
see Table~\ref{orig_stars}). Several (5) of our O-rich AGB stars have 
[5.8]$-$[8.0] vs. [3.6]$-$[4.5] color sufficiently red that they would 
be considered \textit{extreme} AGB stars by \citet{boyer2011}. 
In fact, all of our O-rich AGB stars are redder than the $J-K$ color cut 
\citet{boyer2011} use to isolate O-rich AGB stars in the LMC (see 
Figure~\ref{fig:j-k}). 

\begin{figure}
\includegraphics[trim=0mm 0mm 0mm 17.5mm,angle=90,scale=0.35]{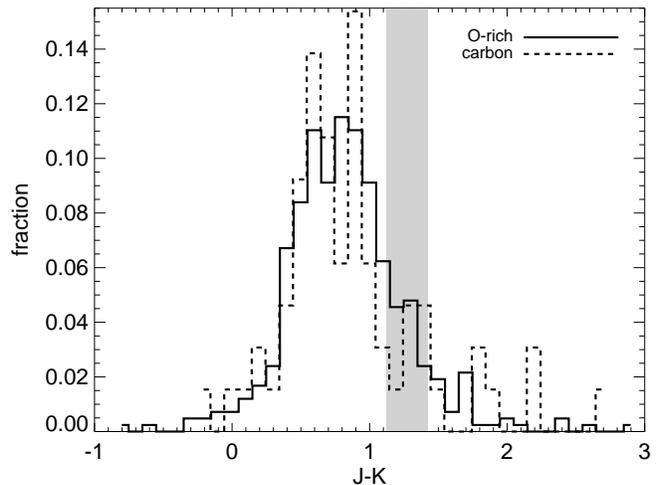}
\caption{Histogram of the $J-K$ color of all AGB stars of known chemical 
  and variability type from the GCVS. $J$ and $K$ photometry from 
  2MASS. The shaded region indicates the $J-K$ color range of the O-rich 
  AGB stars in the \citet{boyer2011} color selection. 
}\label{fig:gcvs_j-k} 
\end{figure}

To test whether the redder colors of our O-rich AGB stars are representative 
of the Galactic AGB population, or rather are the consequence of selection 
effects, we extracted the 2MASS photometry of all Mira and semiregular 
variables in the GCVS \citep{sam12} with known chemical type. The median 
$J-K$ color (0.85 mag) of this larger sample of O-rich 
AGB stars falls within the \citet{boyer2011} color cut, suggesting 
that the red colors of the O-rich AGB stars reflect our choice of O-rich 
AGB stars with measured mass-loss rates, and thus greater IR excess 
(see Figure~\ref{fig:gcvs_j-k}).
It should be noted, however, that a large fraction of the GCVS carbon stars 
have a $J-K$ color falling outside the boundaries of the \citet{boyer2011} 
C-rich color cut. 
\citet{jav11} also find that carbon stars are less red at similar metallicities in M33. 
We find that 78\% of the C-rich AGB stars in the Galaxy are bluer than $J-K \sim 1.2$ mag \citep[the color selection criteria from][for K$\sim -8$ mag, see Figure~\ref{fig:j-k}]{boyer2011}. 
Based on this, the majority of C-rich sources would likely be misclassified as O-rich sources. 
However, sources in the GCVS tend to be optically selected, and as such are 
biased toward low circumstellar extinction (small mass-loss rates), and may sample less-evolved, non-dusty AGB stars that behave very differently than their more evolved, dustier counterparts \citep{boyer2011}. 

\subsection{Source Variability}\label{ss:mag_var}

\begin{figure*}
\includegraphics[angle=90,scale=0.55]{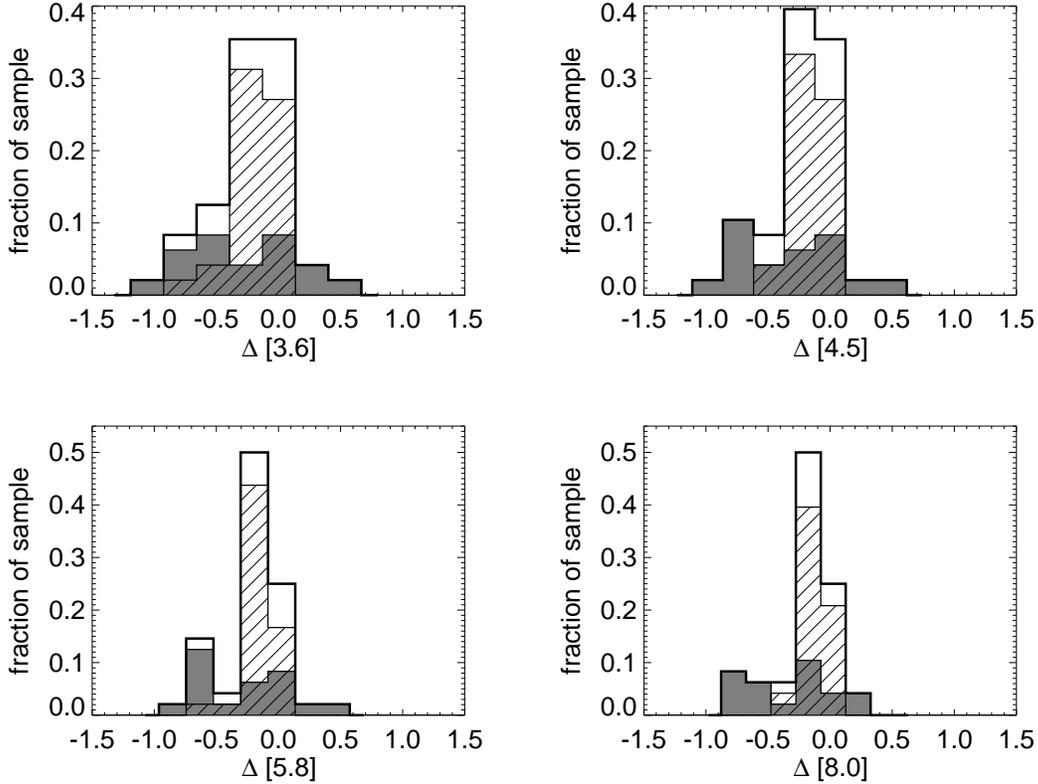}
\caption{Histogram of the magnitude variations between the two epochs
  of observation. %, subtracted by the average magnitude of the sample. 
  White boxes show variations of the full AGB sample, with Mira variables 
  indicated with gray boxes, semiregular and irregulars indicated with 
  striped boxes. }\label{fig:colhist} 
\end{figure*}

The availability of two separate epochs, while not sufficient to
reconstruct a full light-curve, is enough to estimate the average
variations of LPVs in the IRAC bands. Figure~\ref{fig:colhist} shows the 
histograms of the average magnitude change between epochs for all 
sources in our sample. The RMS variations are 0.29, 0.28, 0.27, and 0.26 
mag at 3.6, 4.5, 5.8 and 8.0~\micron\ respectively. This is much larger 
than the RMS variations expected for a sample of non-variable stars 
($\la 0.008$ mag for all bands), based on the average photometric 
uncertainty. As expected, the IRAC bands are affected
by the variability of the sources to a much smaller extent than the
optical bands (where Mira light-curves can have amplitudes as large as
11 mag, \citealt{sam12}), with smaller variations for longer
wavelengths. The smallest changes are observed for sources that are
classified as semiregular or irregular variables 
(0.14, 0.12, 0.11, and 0.10 mag in each band), as expected because
of their shorter periods and smaller pulsation amplitudes with respect
to Miras (0.46, 0.45, 0.42, and 0.41 mag respectively).

The IRAC color changes between epochs are generally small
($\textrm{RMS} \simeq 0.08$ mag), resulting in a small scatter
in the color sequences shown in Figure~\ref{fig:all_color}. The
largest variation is found in the [5.8]$-$[8.0] color for the carbon 
stars in the
blue branch ($\textrm{RMS} \simeq 0.11$ mag), affected by the variability
of the C$_3$ feature (see figure~\ref{fig:carbon_color}). One carbon
star (RS~Cyg) also shows a first epoch [3.6]$-$[8.0] color
significantly bluer (by 0.31 mag) than all other carbon
stars. Unfortunately, ISO SWS spectra are not available for this
source. However, we found that our second epoch IRAC photometry
closely fits the spectrum of another carbon star, T~Ind. The first
epoch IRAC photometry of RS~Cyg is characterised by a much
higher 3.6~\micron\ flux than the second epoch, or the T~Ind spectrum. We
suspect that this fluctuation is due to unusually strong, variable 
C$_2$H$_2$ absorption, common in many of the ISO SWS spectra of the
other carbon stars in our sample. $L$ and $M$ band spectra of RS Cyg
taken at multiple phases would be required to confirm this hypothesis,
and the existence of such broad variations in the strength of this
absorption feature.

%%=============================================================================

\section{Period, Magnitude and Color Relations}\label{s:mag_col_per}

LMC and SMC LPVs tend to organize themselves in a series of parallel
sequences in optical and infrared period-luminosity diagrams (see e.g.
\citealt{w10,riebel2010} and references therein). These sequences
depend on the pulsation mode, binarity, and other not yet
identified parameters (giving rise to poorly understood
characteristics, like a long secondary period observed in some
stars). The so-called \emph{C} sequence is populated by fundamental
mode pulsators. The brightest of the \emph{C} sequence variables tend
to be Miras, while the lower end of the sequence (straddling the RGB
tip) is mainly inhabited by semiregulars. Sequence \emph{C'} (to the
left of sequence \emph{C} by $\Delta [\log P] \simeq - 0.1$), is
instead populated by first overtone semiregular variables. LPVs with
higher overtone modes are organized in separate sequences (\emph{A}
and \emph{B}) with even shorter periods. Stars with the mysterious
secondary long period are found in sequence \emph{D}, shifted by
$\Delta [\log P] \simeq + 0.3$ from the \emph{C} sequence.

\begin{figure}
\includegraphics[trim=0mm 0mm 0mm 17.5mm,angle=90,scale=0.35]{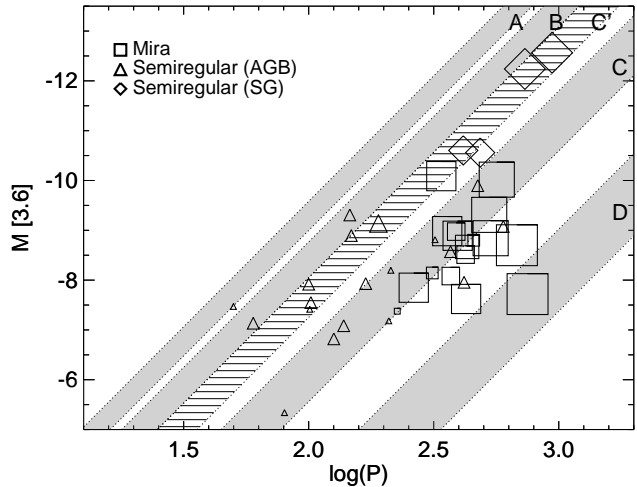}
\caption{3.6~\micron\ absolute magnitude (average of the two epochs)
  versus the logarithm of the period in days for all Miras (squares),
  semiregular AGB variables (triangles), and supergiants
  (diamonds) with a period available in the literature. Shaded and
  striped boxes indicate the approximate distribution of LMC LPVs in
  \citet{riebel2010}. Symbol sizes correspond to [3.6]$-$[8.0] color
  (taken as a proxy for their mass-loss rate).}\label{fig:band1_seq}
\end{figure}

Figure~\ref{fig:band1_seq} shows the period-luminosity diagram 
calculated for the 3.6~\micron\ absolute magnitude (averaged between 
the two epochs) for our sample of AGB stars. Diagrams for the 
other three bands are similar.  The
sources classified in the GCVS \citep{sam12} as Mira and semiregulars
are indicated by different symbols. The size of the symbols is
proportional to the [3.6]$-$[8.0] color (as a proxy for their mass-loss 
rate). The shaded bands for the individual sequences are derived
from the approximate distribution of LMC LPVs in \citet{riebel2010}  
where the [3.6] magnitude is taken as a proxy for the luminosity, 
and converted to absolute magnitudes using the
LMC distance modulus of 18.5 \citep{fre01,wal11}.
These sequences are similar to those found by \citet{w10} using the 
$K$ band magnitude as a proxy for luminosity, however they have a slightly 
shallower slope (larger period for a given luminosity). 

The location of our sources on the diagram is listed in
Table~\ref{tab:PL} and is largely as expected, based on AGB stars in the 
Magellanic Clouds, confirming results found by \citet{glass2009} using 
synthetic IRAC photometry from \citet{mar07}. 
Miras (squares) are all within or below the
sequence \emph{C} (fundamental pulsators), with the exception of the
carbon star R~For, which is in the first overtone sequence \emph{C'}. A few
of the Miras appear to be significantly under-luminous in the 3.6~\micron{}
band, up to 3 mag below the \emph{C} sequence for their period,
placing them in or near the secondary long period sequence \emph{D}. Most of
these stars (e.g. KU~And and V1300~Aql) have large infrared excess due
to their high mass-loss rate; their low 3.6~\micron{} brightness may be
caused by the large extinction due to their thick circumstellar
envelope. 
Alternately, they may have increased their period as the stellar structure 
responds to the mass lost. 
%The case of X~Aqr remains intriguing, as this S-type star is
%two mag fainter than other Miras with similar (312~days) period. 
The semiregular variables, as expected, are mostly located in
the overtone sequences or the lower part of sequence \emph{C}. Only
four semiregulars (T~Ind, VX~And, BN~Mon and RS~Cyg, all carbon stars)
are found in the area of the \emph{C} sequence occupied by fundamental
mode Miras. ET~Vir is the faintest source in the plot, lying at the 
very bottom of sequence \emph{C}. At a Hipparcos-determined distance of 
0.14 kpc, ET~Vir is also the nearest source in the sample. Given the 
large uncertainty on the parallax ($>30$\%), we might expect that the 
luminosity has been underestimated due to an unreliable distance estimate. 
However, the Hipparcos value agrees well with the distance modulus 
calculated using the width of chromospheric Ca~{\sc ii} emission 
lines \citep{wil76} to determine the absolute magnitude, making it 
unclear why ET~Vir is more than a magnitude fainter than the other 
sources in the sample. 
One possibility is that we observed ET~Vir with lower 
luminosity following a thermal pulse. 
All semiregular AGB variables in our sample tend to have
small infrared excess, so it is unlikely that their position in the
diagram is significantly affected by extinction at 3.6~\micron. The
only semiregular variables with large excess are four supergiants
(XX~Per, W~Per, NML~Cyg, and VX~Sgr), 
located at the top of sequence \emph{C'} as expected.

The general distribution of our Galactic LPVs in the period-luminosity
diagram is characterised by a significantly larger spread than the 
variables in the Magellanic Clouds. As found by previous authors (see
e.g. \citealt{tab10}), this is primarily due to the larger uncertainty in
the distance of Galactic LPVs, and possibly in the larger spread in their
metal abundance, which also affects their period-luminosity
relation. Despite these difficulties, the period-luminosity plot using 
the IRAC 3.6~\micron{} band (where extinction is minimal, and infrared
excess tends to be lower than at longer wavelengths) 
as a proxy for luminosity remains an
important tool to study the pulsation mode of these variables, most of
which only possess sparse and incomplete light-curves.

While period-magnitude relations are diagnostics of the pulsation
mechanisms of several classes of variables, the infrared colors of
AGB stars are dominated by dust emission and molecular absorption. The
relationship between infrared colors and period, however, provides an
excellent diagnostic to investigate the dependence of the
circumstellar chemistry and mass-loss rate on the pulsation
period. The [3.6]$-$[8.0] color, which is strongly affected by dust
emission, is a good candidate for this analysis.

\begin{figure}
\includegraphics[trim=0mm 0mm 0mm 17.5mm,angle=90,scale=0.35]{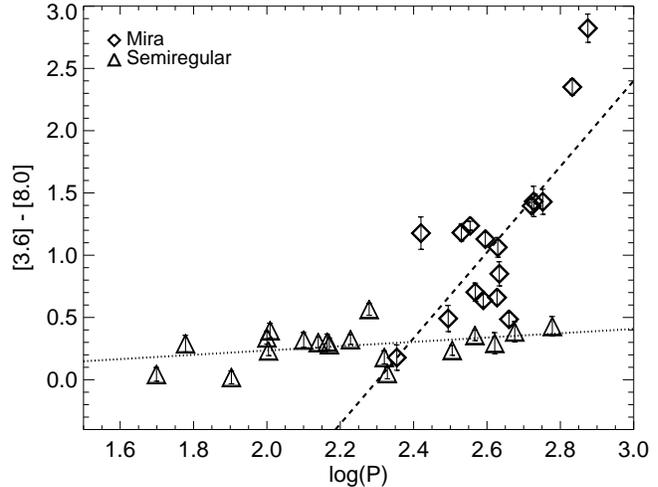}
\caption{IRAC [3.6]$-$[8.0] color (average of the two epochs) versus the
  logarithm of the period in days. A fit to the semiregular variables
  (triangles) shows a strong correlation with period (dotted line). A
  fit to the Miras (squares) is less well-defined (dashed
  line).}\label{fig:period_color}
\end{figure}

The results are shown in figure~\ref{fig:period_color}. Semiregular
variables are distributed along a tight (RMS $\simeq 0.12$ mag)
period-color relation, statistically consistent with a zero slope,
confined within a narrow color range ($\Delta($[3.6]$-$[8.0]$) \la
0.5$ mag). Note that all semiregulars, including the ones with larger period
that appear to be on the fundamental mode sequence \emph{C}, are part
of the same period-color relation. The Mira variables are instead
spread over a much larger range of colors ($\Delta($[3.6]$-$[8.0]$) \approx
3$ mag), with longer period sources (on the \emph{D} sequence)
having the largest infrared excess. The two linear best-fitting relations
are:

\begin{equation}\label{percol_fits}
\left\{
\begin{array}{l}
\mathrm{Semiregular }  \\
\left[3.6]-[8.0] = (-0.11 \pm 0.10) + (0.17 \pm 0.05) \log(P) \right. \\ 
\mathrm{RMS}=0.12 \: \mathrm{mag} \\
\end{array}
\right.
\end{equation}
\begin{equation}\label{percol_fits}
\left\{
\begin{array}{l}
\mathrm{Mira}  \\
\left[3.6]-[8.0] = (-7.93 \pm 0.41) + (3.45 \pm 0.16) \log(P) \right. \\ 
\mathrm{RMS}=0.42 \: \mathrm{mag} \\
\end{array}
\right.
\end{equation}

\noindent
where the period is in days. 

One possible explanation for this dichotomy is
that the two groups of variables are characterised by different dust
composition, affecting their excess in the IRAC bands. One problem
with this interpretation is that there seems to be no difference
between O-rich (M and S-type) and C-rich AGB stars, despite the very
different source of the infrared excess (mainly continuum from 
amorphous carbon dust for carbon
stars, and the 10~\micron\ silicate feature in emission for M and S-type
AGB stars). Another possibility is a difference in the amount of
circumstellar dust responsible for the IR excess, i.e. a different
mass-loss rate. While the median mass-loss rate of the Miras in our 
sample is indeed higher than for the semiregulars 
($2.1 \times 10^{-6}$ and $4.6 \times 10^{-7}$~$M_\odot$~yr$^{-1}$ 
respectively), the separation is not as marked as in the color-period
space. A third possibility is that the separation in color 
reflects different circumstellar dust temperatures. Given
that the IRAC bands are sensitive to dust in the temperature range
$500-800$ K, the observed dichotomy may indicate that semiregulars are
deficient in the hot dust emission that is present in Mira
variables. This is consistent with the suggestion in \citet{mar01}
that semiregular variables could have discontinuous mass-loss,
spending a significant part of their time in the quiescent phase with
only brief phases of higher mass-loss.  According to this hypothesis,
semiregulars would have more excess during these brief intervals of
higher mass-loss, but once this phase has concluded, the dust would
expand and cool, leading to a small [3.6]$-$[8.0] excess ($\la 0.5$ mag). 
Infrared color variations due to intermittent dust production have been found for many AGB stars \citep[see e.g.][and references therein]{whitelock2006}, for both Mira and semiregular LPVs. Those variations have been attributed to the emission of dust clumps, similar to the ones observed with interferometric observations of the very obscured Mira carbon star IRC+10216 \citep{tut2000}. The different color we observe between Mira and SR variables may reflect the fact that dust episodes in semiregulars happen on longer time-scales, giving the dust time to cool between each mass-loss episode.
Less excess for the semiregulars would mean that their IRAC
colors are a better reflection of the star's intrinsic color (and not
thermal emission from the dust). This would lead to the more strict
period-color relation for semiregular than for Mira variables. Indeed it
is the higher excess in Mira variables that would prevent a similarly
tight relation.

%%=============================================================================

\section{Mass Loss and Comparison to LMC/SMC AGB stars}\label{s:comparison}

The intense mass-loss processes that are often active during the AGB phase 
are not well understood. Commonly used empirical mass-loss-rate formulas 
\citep[see e.g.][]{sal74,rei75,baud1983,nie90,vas93} only provide an 
order-of-magnitude estimate of the mass-loss rate of AGB stars and cannot 
predict the actual mass-loss rate as a function of stellar parameters. 
Radiative transfer modeling can be used to infer the amount of circumstellar 
dust from excess flux measurements in the thermal infrared. Coupled with 
assumptions about the AGB wind velocity (usually $\sim 10-20$ km s$^{-1}$) 
and the gas-to-dust mass ratio (typically in the range of the ISM ratio of 
$\sim 200$), these models can be used to provide estimates for AGB mass-loss 
rates from infrared photometric measurements. This can be achieved either by 
fitting the SED \citep[see e.g.][]{gul12,riebel2012} of each source to a 
model or adopting a color -- mass-loss relation 
\citep[see e.g. models using realistic stellar atmospheres from][]{gro06,gro09,gro07,gru08}.
These models show how high-mass-loss rate AGB stars of all types follow 
monotonically increasing sequences similar to those shown in 
Figure~\ref{fig:all_color} (higher mass-loss rate, stronger infrared excess).

\begin{figure}
\includegraphics[trim=0mm 0mm 0mm 17.5mm,angle=90,scale=0.35]{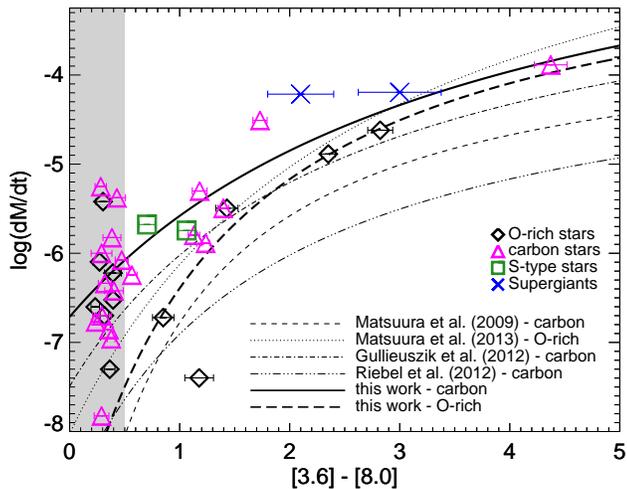}
\caption{A comparison of fits to the mass-loss rate as a function of [3.6]$-$[8.0] 
  color. The best-fitting relations to our sample of Galactic C- and O-rich AGB stars are shown 
  with bold straight and dashed lines respectively. These are compared to 
  fits found by other authors for larger samples of AGB stars in the LMC. 
  Semiregular and irregular LPVs fall within the shaded region 
%  ([3.6]$-$[8.0] < 0.5), 
  while the Mira variables all have redder color.}\label{fig:comp_matsuura}
\end{figure}

Figure~\ref{fig:comp_matsuura} shows the mass-loss rate plotted as function 
of [3.6]$-$[8.0] color. 
Reflecting the color separation found in Figure~\ref{fig:period_color}, 
all semiregular variables are grouped in the $0-0.5$ mag color range. 
However, the overall mass-loss rates are comparable for the two groups. 
This is in accordance with the previously mentioned results found 
in Marengo et al. (2001; see discussion in Section~\ref{s:mag_col_per}), 
and since we would expect semiregular variables to show less excess than 
Mira variables if they are characterised by discontinuous mass-loss. 

The overlap in the distribution of AGB stars of different chemical types echoes 
our finding that the majority of O-rich AGB stars in our sample have colors 
consistent with carbon stars (see Section~\ref{s:colors}). 
Because our sample of O-rich AGB stars is biased towards higher mass-loss rates, 
the most significant contributors of O-rich AGB dust are also the most likely 
to be mis-classified with color cuts alone, suggesting that care must be 
taken in inferring the chemical yield of AGB dust from color-selected 
stellar population studies.

Selecting for AGB stars with measured mass-loss rates may also explain the relatively high mass-loss rate that is characteristic of our sample. 
Our best-fit for the mass-loss rate as a function of [3.6]$-$[8.0] color for the O-rich AGB (excluding the supergiants) and C-rich AGB samples respectively, are: 
\begin{equation}\label{mass_fits}
%\log(\mathrm{dM/dt}) = -72.23 / [([3.6] - [8.0]) + 8.75] - 1.65 
\begin{array}{l}
\mathrm{C-rich:}  \\  
\log(\mathrm{dM/dt}) = -19.20 / [([3.6] - [8.0]) + 3.65] - 1.45 \\
\mathrm{O-rich:}  \\  
%\log(\mathrm{dM/dt}) = -6.82 / [([3.6] - [8.0]) + 0.92] + 2.79 \\
\log(\mathrm{dM/dt}) = -9.65 / [([3.6] - [8.0]) + 1.35] - 2.29 \\
\end{array}
\end{equation}
While the small number of sources in either sample does not allow us to 
establish meaningful uncertainties to the fits, we note that the two 
curves approach each other for $[3.6]-[8.0] \ga 2$ mag. This suggests that 
the mass-loss rate for red sources is similar, for a given value of the 
infrared excess, regardless of the dust chemistry. Our three 
supergiants with reliable mass-loss rate estimates follow a trend 
similar to the AGB stars, but as expected are shifted towards higher mass-loss 
rates.

Similar fits are provided by different authors for AGB stars in the 
Magellanic Clouds. Figure~\ref{fig:comp_matsuura} shows a selection 
of them \citep{matsuura2009,matsuura2013,gul12,riebel2012}, based on the 
photometry obtained as part of the \emph{Spitzer}/IRAC SAGE program. 
It is worth 
noting that the spread between the individual fits is large, as much as 
an order of magnitude in $dM/dt$. This spread may be, in part, a  
consequence of the specific selection criteria adopted by different 
authors, leading to different biases in the AGB samples used in the fit. 
Different assumptions for the gas-to-dust ratio and wind velocity, 
as well as different optical constants used to model the sources will 
also contribute to the spread. While our best-fitting 
relations predict larger mass-loss rates for our Galactic AGB sample, the 
large uncertainty in the fit parameters and the spread between the 
Magellanic Cloud curves prevent us from deriving meaningful conclusions 
about the dependence of mass-loss rates on metallicity. Such a dependence 
is expected for O-rich AGB stars, but not for C-rich sources, since carbon is 
synthesized locally in the Thermally-Pulsing AGB stars, as argued by 
\citet{matsuura2009}. However, it is unclear whether the local synthesis of 
carbon increases the dust-to-gas ratio \citep[see, e.g.][]{van08}. 
A large Galactic sample with reliable distances 
(that could be provided in the near future by GAIA) and radio-based 
mass-loss rate determinations will help resolve this issue.

\begin{figure}%[t!]
\centering
\includegraphics[trim=0mm 0mm 0mm 17.5mm,angle=90,scale=0.35]{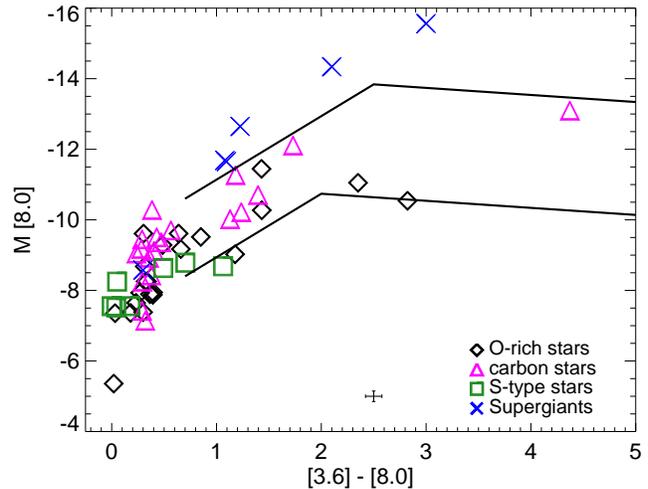}
\caption{Color-Magnitude diagram of IRAC band [8.0]. 
  Average error is plotted in bottom centre. 
  Color-magnitude selection cuts from \citet{matsuura2009} overplotted. 
}\label{fig:color_mag} 
\end{figure}

The selection criteria used by \citet{matsuura2009,matsuura2013} to 
differentiate evolved stars in the Magellanic Clouds result in a sample 
that is more directly comparable with our own Galactic sample. The sources in 
these works are identified by optical and infrared spectroscopy, avoiding 
the potential mis-classification of red O-rich sources (OH/IR stars and 
RSG) in the extreme AGB class (where most of the sources, at least in the 
Magellanic Clouds, are C-rich). \citet{matsuura2013} show how spectrally 
classified sources of different type tend to separate into three 
different regions in the [3.6]$-$[8.0] versus [8.0] color-magnitude diagram. 
These three regions are overlaid on our sample in 
Figure~\ref{fig:color_mag}. 
Note that for our Galactic sources, the bright supergiants 
also have the highest mass-loss rate plotted in 
Figure~\ref{fig:comp_matsuura}. Our two O-rich AGB stars with larger infrared 
excess are in the lowest luminosity region in the \citet{matsuura2013} 
diagram, within the region where extreme C-rich AGB stars are expected. 
However, these two stars are characterised by a mass-loss rate as high as 
C-rich AGB stars with similar excess. As pointed out by \citet{matsuura2013} 
and \citet{riebel2012}, the overall dust return of AGB stars to the ISM 
of the Magellanic Clouds is dominated by a handful of stars with very 
large mass-loss rates. 
In the Magellanic Clouds, most of these stars tend to be C-rich `extreme' 
stars. The situation may be different in the Galaxy where a large number 
of very red OH/IR stars are found. These O-rich evolved stars, with 
mass-loss rates as high as the two reddest O-rich AGB stars in our sample, 
may contribute as much as the C-rich AGB stars to the overall dust 
budget in our Galaxy.  
A larger portion of O-rich AGB stars with high mass-loss rates in sources 
with near solar metallicity has indeed been observed in M33 \citep{jav13}.

\section{Conclusions}\label{s:conclusions}

We present the results of the first study to characterise Galactic AGB 
stars in the IRAC bands. The sample consists of 48 AGB stars observed in 
two epochs -- 22 O-rich, 19 carbon-rich, and 7 S-type -- 
representing a diverse AGB population with well-determined 
distance, chemistry, variability type, and mass-loss rate. These are 
compared to 6 supergiants representing mass-losing evolved stars 
outside the AGB. 

By fitting a PSF to unsaturated parts of the IRAC images, we derive reliable 
photometry in all four IRAC bands. 
From this, we compute the mid-IR colors of Galactic AGB stars. 
AGB stars with O-rich chemistry (including S-type AGB stars) tend to have 
redder [3.6]$-$[8.0] colors than carbon stars for a given [3.6]$-$[4.5] 
color, possibly driven by silicate emission in the [8.0] band. 
For colors including the 5.8 \micron\ band, carbon stars separate into 
two distinct sequences. Carbon stars with higher mass-loss rates tend 
to lie along the redder branch, while sources along the blue branch have 
lower mass-loss rates. We interpret this as evidence of a photospheric 
C$_3$ feature that is only visible in the low-mass-loss-rate sources 
that are relatively unobscured by dusty circumstellar envelopes. 

AGB stars are LPVs, so we have examined both the color and the magnitude 
as a function of period. 
We find a period-color relationship consistent with the hypothesis 
of \citet{mar01} that semiregular variables lose mass discontinuously, 
leading to a lack of hot dust producing redder colors. In contrast, 
the Mira variables extend along a more linear sequence toward redder 
colors as we might expect based on the fact that their mass-loss rates 
are more sustained in time. 

The period-luminosity distribution of Galactic AGB stars is similar to 
that found by \citet{riebel2012} for the LMC. 
Mira variables fall along the fundamental pulsation sequence \emph{C}.  
Semiregular variables are mostly in sequences \emph{A} and \emph{B}, 
reflecting the presence of higher overtone modes and shorter periods. 

We derive a relationship between mass-loss rate and [3.6]$-$[8.0] color 
separately for O-rich AGB stars and carbon stars. 
The overall shape of the fits is similar to those found by other authors 
for AGB stars in the LMC, albeit corresponding to a higher mass-loss rate 
for a given [3.6]$-$[8.0] color. Discrepancies between our fits and those 
found by other authors likely reflect different assumptions used in the 
radiative transfer models used to derive mass-loss rates and uncertainties 
in the gas-to-dust ratio. In addition, we find that neither 
color nor mass-loss rate is a good discriminator of chemical type, 
suggesting that care must be taken when inferring the chemical 
contribution of dust returned to the ISM by AGB stars.

%%=============================================================================

\section*{Acknowledgements}
We thank the referee, Jacco van Loon, for a thoughtful review 
and suggestions that improved the quality of the manuscript. 
This work is based on observations made with the \emph{Spitzer} Space
Telescope, which is operated by the Jet Propulsion Laboratory,
California Institute of Technology under NASA contract 1407. 
This research has made use of the SIMBAD database, operated at CDS, 
Strasbourg, France. Support
for this work was provided by NASA through an award issued by
JPL/Caltech.  This work is also supported by the National Science
Foundation's Research Experience for Undergraduates program. MR 
wishes to thank the REU coordinators Saku Vrtilek, Christine
Jones, Melissa Cirtain, and Jonathan McDowell for all their guidance
and support.

%{\it Facilities:} \facility{Spitzer (IRAC)}, \facility{2MASS}

%%=============================================================================

%\clearpage

%%=============================================================================

%%% A whole lot of tables

\begin{table*}
\caption{IRAC AGB OBSERVED TARGETS}
\begin{tabular}{lrlrrrcrcrr}
\hline\hline
Target & IRAS name & Var. & Period & Dist & Dist & $\dot{\mathrm{M}}$ & $\dot{\mathrm{M}}$ & J Mag & H Mag & K Mag \\
 &  &  & [days] & [kpc] & ref. & [M$_\odot$/yr] & ref. &  &  &  \\ 
\hline
\multicolumn{11}{c}{O-RICH STARS} \\
\hline
KU And & 00042$+$4248 & M & 750 & 0.68 & 14 & 2.4$\times 10^{-5}$ & 11 & 3.041$\pm$0.216 & 1.829$\pm$0.156 & 1.115$\pm$0.208 \\
RW And & 00445$+$3224 & M & 430 & 0.86 & 7 & 1.9$\times 10^{-7}$ & 17 & 3.052$\pm$0.238 & 2.225$\pm$0.210 & 1.765$\pm$0.228 \\
VY Cas & 00484$+$6238 & SRb & 100 & 0.57 & 16 & ... & ... & 2.471$\pm$0.254 & 1.484$\pm$0.186 & 1.114$\pm$0.234 \\
SV Psc & 01438$+$1850 & SRb & 102 & 0.38 & 12 & 3.0$\times 10^{-7}\dagger$ & 12 & 2.016$\pm$0.244 & 1.012$\pm$0.190 & 0.722$\pm$0.188 \\
RR Per & 02251$+$5102 & M & 389 & 0.82 & 15 & ... & ... & 3.217$\pm$0.244 & 2.325$\pm$0.196 & 1.662$\pm$0.244 \\ 
RV Cam & 04265$+$5718 & SRb & 101 & 0.35 & 12 & 2.5$\times 10^{-7}$ & 12 & 1.667$\pm$0.254 & 0.602$\pm$0.164 & 0.412$\pm$0.170 \\
ET Vir & 14081$-$1604 & SRb & 80 & 0.14 & 16 & ... & ... & 1.651$\pm$0.266 & 0.814$\pm$0.290 & 0.598$\pm$0.304 \\
RW Boo & 14390$+$3147 & SRb & 209 & 0.29 & 16 & ... & ... & 1.560$\pm$0.206 & 0.530$\pm$0.150 & 0.301$\pm$0.170 \\
AX Sco & 16387$-$2700 & SRb & 138 & 0.34 & 16 & ... & ... & 2.159$\pm$0.270 & 1.224$\pm$0.282 & 0.901$\pm$0.320 \\
TV Dra & 17081$+$6422 & SR & ... & 0.51 & 16 & 2.0$\times 10^{-7}$ & 11 & 1.942$\pm$0.416 & 1.058$\pm$0.174 & 0.693$\pm$0.194 \\
V438 Oph & 17123$+$1107 & SRb & 169 & 0.42 & 4 & ... & ... & 1.871$\pm$0.244 & 0.935$\pm$0.292 & 0.551$\pm$0.312 \\
TY Dra & 17361$+$5746 & Lb & ... & 0.43 & 12 & 6.0$\times 10^{-7}$ & 12 & 2.410$\pm$0.296 & 1.465$\pm$0.158 & 1.083$\pm$0.194 \\
CZ Ser & 18347$-$0241 & Lb & ... & 0.44 & 12 & 8.0$\times 10^{-7}$ & 12 & 2.557$\pm$0.302 & 1.405$\pm$0.274 & 0.930$\pm$0.310 \\
FI Lyr & 18401$+$2854 & SRb & 146 & 0.88 & 6 & 3.8$\times 10^{-6}$ & 6 & 2.040$\pm$0.282 & 1.121$\pm$0.164 & 0.715$\pm$0.212 \\
V1351 Cyg & 19409$+$5520 & Lb & ... & 0.32 & 16 & ... & ... & 1.816$\pm$0.266 & 0.787$\pm$0.170 & 0.561$\pm$0.208 \\
Z Cyg & 20000$+$4954 & M & 263 & 0.94 & 5 & 4.0$\times 10^{-8}$ & 19 & 4.176$\pm$0.268 & 3.258$\pm$0.228 & 2.557$\pm$0.306 \\
V1300 Aql & 20077$-$0625 & M & 680 & 0.66 & 11 & 1.3$\times 10^{-5}$ & 11 & 6.906$\pm$0.032 & 3.923$\pm$0.260 & 2.059$\pm$0.262 \\
V584 Aql & 20079$-$0146 & Lb & ... & 0.39 & 12 & 5.0$\times 10^{-8}$ & 12 & 2.136$\pm$0.312 & 1.147$\pm$0.204 & 0.801$\pm$0.266 \\
RX Vul & 20507$+$2310 & M & 457 & 0.71 & 18 & ... & ... & 2.724$\pm$0.290 & 1.655$\pm$0.188 & 1.083$\pm$0.228\\
UX Cyg & 20529$+$3013 & M & 565 & 1.85 & 9 & 3.2$\times 10^{-6}$ & 11 & 2.910$\pm$0.310 & 1.887$\pm$0.180 & 1.400$\pm$0.194\\
SS Peg & 22315$+$2418 & M & 424 & 0.71 & 18 & ... & ... & 2.549$\pm$0.270 & 1.513$\pm$0.182 & 1.121$\pm$0.162 \\
V563 Cas & 23147$+$6009 & M & 534 & 2.09 & 11 & ... & ... & 5.088$\pm$0.037 & 3.882$\pm$0.036 & 3.227$\pm$0.248 \\
\hline
\multicolumn{11}{c}{C-RICH STARS} \\
\hline
VX And & 00172$+$4425 & SRa & 369 & 0.56 & 6 & 1.4$\times 10^{-7}$ & 6 & 3.187$\pm$0.266 & 1.891$\pm$0.202 & 1.193$\pm$0.202\\

HV Cas & 01080$+$5327 & M & 527 & 0.97 & 14 & 3.2$\times 10^{-6}$ & 6 & 5.585$\pm$0.039 & 3.911$\pm$0.222 & 2.466$\pm$0.290 \\

R For & 02270$-$2619 & M & 339 & 0.97 & 14 & 5.0$\times 10^{-6}$ & 6 & 
4.230$\pm$0.274 & 2.537$\pm$0.206 & 1.349$\pm$0.274 \\

V623 Cas & 03075$+$5742 & Lb & ... & 0.51 & 6 & 1.1$\times 10^{-7}$ & 6 & 2.869$\pm$0.210 & 1.706$\pm$0.204 & 1.208$\pm$0.212 \\

UV Aur & 05185$+$3227 & M & 394 & 1.09 & 6 & 1.6$\times 10^{-6}$ & 6 & 4.029$\pm$0.214 & 3.018$\pm$0.192 & 2.129$\pm$0.220 \\

SY Per & 04127$+$4030 & SR & 474 & 1.43 & 6 & 1.5$\times 10^{-6}$ & 6 & 4.596$\pm$0.316 & 3.033$\pm$0.274 & 2.132$\pm$0.340 \\ 

TU Tau & 05421$+$2424 & SRb & 190 & 1.05 & 6 & 5.7$\times 10^{-7}$ & 6 & 3.331$\pm$0.266 & 2.093$\pm$0.210 & 1.574$\pm$0.220 \\

BN Mon & 06192$+$0722 & SRb & 600 & 1.28 & 1 & 4.2$\times 10^{-6}$ & 6 & 4.517$\pm$0.312 & 3.096$\pm$0.256 & 2.322$\pm$0.308 \\

CR Gem & 06315$+$1606 & Lb & ... & 0.92 & 6 & 8.4$\times 10^{-7}$ & 6 & 3.575$\pm$0.302 & 2.030$\pm$0.250 & 1.538$\pm$0.240 \\

V614 Mon & 06585$-$0310 & SRb & 60 & 0.48 & 6 & 1.5$\times 10^{-8}$ & 6 & 
3.227$\pm$0.206 & 2.307$\pm$0.206 & 1.764$\pm$0.224 \\

CGCS 6296 & 08305$-$3314 & ... & ... & 4.21 & 6 & 1.3$\times 10^{-4}$ & 6 & ... & ... & 11.1 \\

SZ Car & 09582$-$5958 & SRb & 126 & 0.37 & 6 & 4.6$\times 10^{-7}$ & 6 & 3.206$\pm$0.332 & 2.112$\pm$0.280 & 1.600$\pm$0.346 \\

V CrB & 15477$+$3943 & M & 358 & 0.84 & 6 & 1.3$\times 10^{-6}$ & 6 & 3.474$\pm$0.272 & 2.209$\pm$0.218 & 1.321$\pm$0.276 \\

SX Sco & 17441$-$3541 & SR & ... & 0.83 & 6 & 3.8$\times 10^{-7}$ & 6 & 3.104$\pm$0.222 & 1.884$\pm$0.210 & 1.390$\pm$0.254 \\

FX Ser & 18040$-$0941 & Lb & ... & 1.23 & 6 & 3.1$\times 10^{-5}$ & 6 & 7.136$\pm$0.023 & 4.720$\pm$0.076 & 2.702$\pm$0.218 \\

DR Ser & 18448$+$0523 & Lb & ... & 1.29 & 6 & 1.0$\times 10^{-6}$ & 6 & 3.833$\pm$0.270 & 2.488$\pm$0.218 & 1.839$\pm$0.286 \\

S Sct & 18476$-$0758 & SRb & 148 & 0.58 & 6 & 5.6$\times 10^{-6}$ & 6 & 
2.303$\pm$0.314 & 1.140$\pm$0.262 & 0.627$\pm$0.288 \\

RS Cyg & 20115$+$3834 & SRa & 418 & 0.65 & 6 & 2.0$\times 10^{-7}$ & 6 & 3.206$\pm$0.228 & 1.985$\pm$0.190 & 1.321$\pm$0.246 \\

T Ind & 21168$-$4514 & SRb & 320 & 0.65 & 6 & 1.7$\times 10^{-7}$ & 6 & 1.981$\pm$0.248 & 0.970$\pm$0.262 & 0.564$\pm$0.316 \\
\hline
\multicolumn{11}{c}{S-RICH STARS} \\
\hline
R Gem & 07043$+$2246 & M & 370 & 0.71 & 13 & 2.1$\times 10^{-6}$ & 11 & 2.530$\pm$0.244 & 1.642$\pm$0.218 & 1.459$\pm$0.226 \\

NQ Pup & 07507$-$1129 & Lb & ... & 0.81 & 16 & ... & ... & 3.542$\pm$0.292 & 2.587$\pm$0.212 & 2.306$\pm$0.274 \\

S UMa & 12417$+$6121 & M & 226 & 1.09 & 18 & ... & ... & 4.458$\pm$0.214 & 3.431$\pm$0.196 & 3.017$\pm$0.246 \\

R Cyg & 19354$+$5005 & M & 426 & 0.44 & 14 & 1.6$\times 10^{-7}$ & 11 & 2.251$\pm$0.314 & 1.379$\pm$0.208 & 0.861$\pm$0.246 \\

AA Cyg & 20026$+$3640 & SRb & 213 & 0.48 & 13 & ... & ... & 2.066$\pm$0.288 & 1.050$\pm$0.166 & 0.625$\pm$0.224 \\

X Aqr & 22159$-$2109 & M & 312 & 1.32 & 18 & ... & ... & 4.710$\pm$0.190 & 3.823$\pm$0.196 & 3.338$\pm$0.248 \\

HR Peg & 22521$+$1640 & SRb & 50 & 0.42 & 16 & ... & ... & 2.306$\pm$0.270 & 1.239$\pm$0.194 & 1.041$\pm$0.206 \\
\hline
\multicolumn{11}{c}{SUPERGIANTS} \\
\hline
XX Per & 01597$+$5459 & SRc & 415 & 2.29 & 3 & ... & ... & 3.442$\pm$0.262
& 2.480$\pm$0.232 & 1.972$\pm$0.242 \\

W Per & 02469$+$5646 & SRc & 485 & 2.29 & 3 & ... & ... & 3.095$\pm$0.202
& 1.999$\pm$0.172 & 1.568$\pm$0.252 \\

NO Aur & 05374$+$3153 & Lc & ... & 0.60 & 16 & ... & ... & 2.122$\pm$0.322 & 1.128$\pm$0.202 & 0.971$\pm$0.196\\

U Lac & 22456$+$5453 & SRc & ... & 2.75 & 10 & ... & ... & 3.815$\pm$0.224
& 2.672$\pm$0.166 & 1.932$\pm$0.206 \\

NML Cyg$*$ & R20445$+$3955 & SRc & 940$\ddagger$ & 1.61 & 20 & 6.4$\times 10^{-5}$ & 21,22 & 
4.877$\pm$0.037 & 2.389$\pm$0.200 & 0.791$\pm$0.204 \\

VX Sgr$*$ & 18050$-$2213 & SRc & 732 & 1.57 & 2 & 
6.1$\times 10^{-5}$ & 4 & 
1.744$\pm$0.260 & 0.550$\pm$0.304 & $-$0.122$\pm$0.362 \\
\hline
\end{tabular}
\label{orig_stars}
\medskip
\footnotesize
\vspace{-10pt}
\begin{tabular}{l}
$*$ from Table 4.4 in \citet{sch07}, $\dagger$ multiple velocity components detected, $\ddagger$ \citealt{mon97}. \\

\textbf{References:} 
(1) \citealt{ber05}; 
(2) \citealt{che07}; 
(3) \citealt{cur10}; 
(4) \citealt{debeck2010}; 
(5) \citealt{fea89}; \\
(6) \citealt{gua06}; 
(7) \citealt{gua08}; 
(8) \citealt{hum78}; 
(9) \citealt{kur05}; 
(10) \citealt{lev05}; \\
(11) \citealt{lou93}; 
(12) \citealt{olo02}; 
(13) \citealt{ram09}; 
(14) \citealt{sch13}; 
(15) \citealt{van02}; \\
(16) \citealt{van07}; 
(17) \citealt{whitelock1994}; 
(18) \citealt{whitelock2008}; 
%(19) \citealt{win03}; 
(19) \citealt{you95}; 
(20) \citealt{zha12}; \\
(21) \citealt{hyl72}; and 
(22) \citealt{mor83}. 
\end{tabular}
\end{table*}

\clearpage

%%=============================================================================

%%%IRAC photometry

\begin{table*}
\caption{IRAC PHOTOMETRY}
\centering
\begin{tabular}{lrcrrrrr}
\hline\hline
Target & IRAS name & epoch & MJD$^a$ & [3.6] & [4.5] & [5.8] & [8.0] \\ 
\hline
%%\cutinhead{O-RICH STARS}
\multicolumn{8}{c}{O-RICH STARS} \\
\hline

KU And  &  00042$+$4248  &  1  &  53961.9209  &  1.140$\pm$0.031  &  0.291$\pm$0.021  &  $-$0.400$\pm$0.015  &  $-$1.584$\pm$0.076  \\ 
KU And  &  00042$+$4248  &  2  &  54328.9385  &  1.747$\pm$0.027  &  0.879$\pm$0.024  &  0.108$\pm$0.018  &  $-$1.175$\pm$0.074  \\ 

RW And  &  00445$+$3224  &  1  &  53960.9362  &  1.109$\pm$0.030  &  1.042$\pm$0.028  &  0.742$\pm$0.022  &  0.216$\pm$0.066  \\
RW And  &  00445$+$3224  &  2  &  54324.8287  &  0.904$\pm$0.025  &  0.831$\pm$0.023  &  0.546$\pm$0.027  &  0.096$\pm$0.059  \\

VY Cas  &  00484$+$6238  &  1  &  54005.9215  &  0.879$\pm$0.024  &  0.981$\pm$0.027  &  0.786$\pm$0.022  &  0.564$\pm$0.027  \\ 
VY Cas  &  00484$+$6238  &  2  &  54149.6210  &  0.843$\pm$0.024  &  0.929$\pm$0.026  &  0.731$\pm$0.032  &  0.485$\pm$0.025  \\ 

SV Psc  &  01438$+$1850  &  1  &  53961.9248  &  0.395$\pm$0.023  &  0.468$\pm$0.025  &  0.256$\pm$0.021  &  $-$0.027$\pm$0.042  \\ 
SV Psc  &  01438$+$1850  &  2  &  54149.6256  &  0.305$\pm$0.015  &  0.411$\pm$0.016  &  0.202$\pm$0.020  &  $-$0.063$\pm$0.031  \\

RR Per  &  02251$+$5102  &  1  &  54005.9269  &  0.427$\pm$0.024  &  0.320$\pm$0.015  &  0.044$\pm$0.017  &  $-$0.289$\pm$0.033  \\ 
RR Per  &  02251$+$5102  &  2  &  54149.9993  &  0.764$\pm$0.022  &  0.775$\pm$0.022  &  0.502$\pm$0.017  &  0.202$\pm$0.020  \\

RV Cam  &  04265$+$5718  &  1  &  54005.9347  &  0.327$\pm$0.015  &  0.451$\pm$0.016  &  0.277$\pm$0.021  &  0.073$\pm$0.023  \\ 
RV Cam  &  04265$+$5718  &  2  &  54396.8337  &  0.284$\pm$0.014  &  0.435$\pm$0.016  &  0.216$\pm$0.020  &  0.073$\pm$0.023  \\

ET Vir  &  14081$-$1604  &  1  &  53957.3504  &  0.395$\pm$0.016  &  0.573$\pm$0.018  &  0.443$\pm$0.024  &  0.372$\pm$0.031  \\ 
ET Vir  &  14081$-$1604  &  2  &  54148.6245  &  0.387$\pm$0.023  &  0.537$\pm$0.027  &  0.427$\pm$0.024  &  0.372$\pm$0.031  \\

RW Boo  &  14390$+$3147  &  1  &  53922.4493  &  0.127$\pm$0.018  &  0.387$\pm$0.016  &  0.145$\pm$0.019  &  $-$0.053$\pm$0.032  \\
RW Boo  &  14390$+$3147  &  2  &  54149.9923  &  0.133$\pm$0.018  &  0.387$\pm$0.016  &  0.170$\pm$0.019  &  $-$0.048$\pm$0.033  \\

AX Sco  &  16387$-$2700  &  1  &  53997.8064  &  0.564$\pm$0.018  &  0.649$\pm$0.020  &  0.435$\pm$0.016  &  0.256$\pm$0.021  \\
AX Sco  &  16387$-$2700  &  2  &  54357.9155  &  0.582$\pm$0.019  &  0.700$\pm$0.015  &  0.485$\pm$0.017  &  0.291$\pm$0.028  \\

TV Dra  &  17081$+$6422  &  1  &  54005.9137  &  0.177$\pm$0.021  &  0.263$\pm$0.021  &  0.061$\pm$0.017  &  $-$0.123$\pm$0.019  \\ 
TV Dra  &  17081$+$6422  &  2  &  54286.3724  &  0.170$\pm$0.019  &  0.242$\pm$0.020  &  0.022$\pm$0.017  &  $-$0.161$\pm$0.019  \\ 

V438 Oph  &  17123$+$1107  &  1  &  54002.9634  &  0.209$\pm$0.020  &  0.294$\pm$0.014  &  0.096$\pm$0.018  &  $-$0.094$\pm$0.020  \\ 
V438 Oph  &  17123$+$1107  &  2  &  54357.9197  &  0.164$\pm$0.019  &  0.263$\pm$0.021  &  0.073$\pm$0.017  &  $-$0.180$\pm$0.020  \\ 

TY Dra  &  17361$+$5746  &  1  &  54005.9110  &  0.679$\pm$0.020  &  0.891$\pm$0.020  &  0.649$\pm$0.030  &  0.284$\pm$0.042  \\ 
TY Dra  &  17361$+$5746  &  2  &  54286.3697  &  0.689$\pm$0.020  &  0.879$\pm$0.024  &  0.659$\pm$0.030  &  0.298$\pm$0.057  \\  

CZ Ser  &  18347$-$0241  &  1  &  54005.8978  &  0.519$\pm$0.026  &  0.710$\pm$0.021  &  0.435$\pm$0.024  &  0.277$\pm$0.028  \\ 
CZ Ser  &  18347$-$0241  &  2  &  54229.6547  &  0.591$\pm$0.037  &  0.742$\pm$0.032  &  0.476$\pm$0.025  &  0.291$\pm$0.028  \\ 

FI Lyr  &  18401$+$2854  &  1  &  54005.9018  &  0.411$\pm$0.016  &  0.601$\pm$0.028  &  0.357$\pm$0.023  &  0.108$\pm$0.030  \\ 
FI Lyr  &  18401$+$2854  &  2  &  54357.9236  &  0.419$\pm$0.024  &  0.601$\pm$0.019  &  0.372$\pm$0.023  &  0.114$\pm$0.048  \\

V1351 Cyg  &  19409$+$5520  &  1  &  53956.9174  &  0.196$\pm$0.020  &  0.395$\pm$0.023  &  0.229$\pm$0.020  &  0.177$\pm$0.019  \\ 
V1351 Cyg  &  19409$+$5520  &  2  &  54286.3181  &  0.209$\pm$0.020  &  0.403$\pm$0.016  &  0.222$\pm$0.020  &  0.164$\pm$0.025  \\ 

Z Cyg  &  20000$+$4954  &  1  &  54005.9076  &  2.174$\pm$0.056  &  1.892$\pm$0.043  &  1.620$\pm$0.047  &  0.929$\pm$0.077  \\
Z Cyg  &  20000$+$4954  &  2  &  54286.3642  &  1.862$\pm$0.060  &  1.669$\pm$0.051  &  1.422$\pm$0.060  &  0.753$\pm$0.065  \\ 

V1300 Aql  &  20077$-$0625  &  1  &  54430.9969  &  0.145$\pm$0.025  &  $-$0.483$\pm$0.028  &  $-$1.187$\pm$0.022  &  $-$2.280$\pm$0.040  \\ 
V1300 Aql  &  20077$-$0625  &  2  &  54637.7548  &  0.649$\pm$0.020  &  0.050$\pm$0.017  &  $-$0.647$\pm$0.024  &  $-$1.627$\pm$0.036  \\ 

V584 Aql  &  20079$-$0146  &  1  &  54064.4277  &  0.427$\pm$0.016  &  0.502$\pm$0.017  &  0.305$\pm$0.022  &  0.091$\pm$0.024  \\ 
V584 Aql  &  20079$-$0146  &  2  &  54430.9864  &  0.427$\pm$0.016  &  0.502$\pm$0.017  &  0.294$\pm$0.021  &  0.033$\pm$0.034  \\ 

RX Vul  &  20507$+$2310  &  1  &  54286.3354  &  0.419$\pm$0.024  &  0.468$\pm$0.017  &  0.170$\pm$0.019  &  $-$0.104$\pm$0.020  \\
RX Vul  &  20507$+$2310  &  2  &  54430.9901  &  0.493$\pm$0.026  &  0.610$\pm$0.029  &  0.342$\pm$0.022  &  0.044$\pm$0.034  \\

UX Cyg  &  20529$+$3013  &  1  &  54286.3327  &  1.009$\pm$0.027  &  0.731$\pm$0.021  &  0.411$\pm$0.024  &  $-$0.403$\pm$0.075  \\ 
UX Cyg  &  20529$+$3013  &  2  &  54430.9927  &  1.629$\pm$0.034  &  1.307$\pm$0.036  &  0.968$\pm$0.026  &  0.183$\pm$0.051  \\ 

SS Peg  &  22315$+$2418  &  1  &  54286.3777  &  0.820$\pm$0.023  &  0.649$\pm$0.020  &  0.380$\pm$0.023  &  0.005$\pm$0.022  \\ 
SS Peg  &  22315$+$2418  &  2  &  54100.2776  &  0.669$\pm$0.030  &  0.669$\pm$0.020  &  0.419$\pm$0.016  &  0.164$\pm$0.038  \\

V563 Cas  &  23147$+$6009  &  1  &  53956.1633  &  2.585$\pm$0.035  &  2.097$\pm$0.037  &  1.684$\pm$0.041  &  1.036$\pm$0.056  \\ 
V563 Cas  &  23147$+$6009  &  2  &  54149.4121  &  2.935$\pm$0.032  &  2.639$\pm$0.037  &  2.190$\pm$0.041  &  1.620$\pm$0.097  \\ 

%%
%\cutinhead{C-RICH STARS}
\hline
\multicolumn{8}{c}{C-RICH STARS} \\
\hline

VX And  &  00172$+$4425  &  1  &  53961.0086  &  0.158$\pm$0.019  &  0.357$\pm$0.023  &  0.485$\pm$0.025  &  $-$0.216$\pm$0.027  \\ 
VX And  &  00172$+$4425  &  2  &  54150.6613  &  0.183$\pm$0.019  &  0.395$\pm$0.023  &  0.528$\pm$0.018  &  $-$0.157$\pm$0.019  \\ 

HV Cas  &  01080$+$5327  &  1  &  53961.4766  &  0.891$\pm$0.025  &  0.342$\pm$0.015  &  $-$0.108$\pm$0.029  &  $-$0.589$\pm$0.032  \\ 
HV Cas  &  01080$+$5327  &  2  &  54150.6238  &  0.357$\pm$0.015  &  $-$0.152$\pm$0.019  &  $-$0.573$\pm$0.016  &  $-$0.955$\pm$0.023  \\ 

R For   &  02270$-$2619  &  1  &  53961.0145  &  $-$0.476$\pm$0.018  &  $-$0.900$\pm$0.019  &  $-$1.227$\pm$0.018  &  $-$1.621$\pm$0.037  \\
R For   &  02270$-$2619  &  2  &  54357.9070  &  $-$0.354$\pm$0.016  &  $-$0.811$\pm$0.016  &  $-$1.160$\pm$0.019  &  $-$1.571$\pm$0.051  \\

V623 Cas  &  03075$+$5742  &  1  &  54005.9318  &  0.485$\pm$0.017  &  0.546$\pm$0.018  &  0.305$\pm$0.022  &  0.102$\pm$0.018  \\ 
V623 Cas  &  03075$+$5742  &  2  &  54357.9005  &  0.493$\pm$0.017  &  0.564$\pm$0.018  &  0.357$\pm$0.023  &  0.120$\pm$0.018  \\ 

SY Per  &  04127$+$4030  &  1  &  54396.8364  &  0.820$\pm$0.046  &  0.786$\pm$0.034  &  0.649$\pm$0.040  &  0.605$\pm$0.029  \\ 
SY Per  &  04127$+$4030  &  2  &  54531.1417  &  0.942$\pm$0.039  &  0.879$\pm$0.037  &  0.775$\pm$0.033  &  0.387$\pm$0.047  \\ 

UV Aur  &  05185$+$3227  &  1  &  54005.9383  &  1.422$\pm$0.029  &  1.017$\pm$0.022  &  0.620$\pm$0.019  &  0.164$\pm$0.025  \\ 
UV Aur  &  05185$+$3227  &  2  &  54191.1691  &  1.171$\pm$0.026  &  0.855$\pm$0.024  &  0.451$\pm$0.025  &  0.170$\pm$0.025  \\ 

TU Tau  &  05421$+$2424  &  1  &  54396.8428  &  0.981$\pm$0.027  &  1.094$\pm$0.029  &  0.929$\pm$0.038  &  0.411$\pm$0.024  \\ 
TU Tau  &  05421$+$2424  &  2  &  54188.7098  &  0.955$\pm$0.021  &  1.051$\pm$0.029  &  0.916$\pm$0.025  &  0.395$\pm$0.023  \\ 

BN Mon  &  06192$+$0722  &  1  &  54038.7664  &  1.463$\pm$0.029  &  1.550$\pm$0.036  &  1.422$\pm$0.040  &  1.036$\pm$0.042  \\
BN Mon  &  06192$+$0722  &  2  &  54396.8610  &  1.454$\pm$0.021  &  1.550$\pm$0.032  &  1.463$\pm$0.042  &  1.022$\pm$0.055  \\ 
\hline
\end{tabular}
\label{phot_table}
\end{table*}

\begin{table*}
\contcaption{}
\centering
\begin{tabular}{lrcrrrrr}
\hline\hline
Target & IRAS name & epoch & MJD$^a$ & [3.6] & [4.5] & [5.8] & [8.0] \\ 
\hline
\multicolumn{8}{c}{C-RICH STARS} \\
\hline
CR Gem  &  06315$+$1606  &  1  &  54061.8204  &  0.995$\pm$0.027  &  0.916$\pm$0.025  &  0.601$\pm$0.028  &  0.485$\pm$0.034  \\ 
CR Gem  &  06315$+$1606  &  2  &  54396.8582  &  0.855$\pm$0.024  &  0.867$\pm$0.024  &  0.555$\pm$0.027  &  0.419$\pm$0.032  \\ 

V614 Mon  &  06585$-$0310  &  1  &  54069.0020  &  1.272$\pm$0.025  &  1.422$\pm$0.028  &  1.254$\pm$0.026  &  0.955$\pm$0.039  \\ 
V614 Mon  &  06585$-$0310  &  2  &  54190.6094  &  1.272$\pm$0.025  &  1.422$\pm$0.032  &  1.265$\pm$0.028  &  1.009$\pm$0.041  \\ 

CGCS 6296  &  08305$-$3314  &  1  &  54466.0597  &  4.120$\pm$0.048  &  2.447$\pm$0.052  &  1.124$\pm$0.031  &  $-$0.133$\pm$0.029  \\ 
CGCS 6296  &  08305$-$3314  &  2  &  54600.7407  &  4.674$\pm$0.121  &  2.935$\pm$0.049  &  1.550$\pm$0.045  &  0.183$\pm$0.064  \\ 

SZ Car  &  09582$-$5958  &  1  &  53888.8675  &  1.065$\pm$0.029  &  1.171$\pm$0.032  &  1.065$\pm$0.029  &  0.742$\pm$0.032  \\ 
SZ Car  &  09582$-$5958  &  2  &  54148.6298  &  0.976$\pm$0.027  &  1.080$\pm$0.029  &  0.916$\pm$0.038  &  0.659$\pm$0.030  \\ 

V CrB  &  15477$+$3943  &  1  &  53959.8338  &  0.874$\pm$0.019  &  0.387$\pm$0.016  &  $-$0.048$\pm$0.021  &  $-$0.385$\pm$0.015  \\ 
V CrB  &  15477$+$3943  &  2  &  54152.9619  &  0.403$\pm$0.016  &  $-$0.063$\pm$0.015  &  $-$0.447$\pm$0.014  &  $-$0.811$\pm$0.021  \\ 

SX Sco &  17441$-$3541  &  1  &  53997.2559  &  0.891$\pm$0.049  &  1.022$\pm$0.042  &  0.942$\pm$0.039  &  0.476$\pm$0.034  \\ 
SX Sco &  17441$-$3541  &  2  &  54396.9331  &  0.867$\pm$0.048  &  1.065$\pm$0.043  &  0.995$\pm$0.041  &  0.476$\pm$0.034  \\ 

FX Ser &  18040$-$0941  &  1  &  54002.9673  &  $-$0.118$\pm$0.019  &  $-$0.787$\pm$0.016  &  $-$1.298$\pm$0.023  &  $-$1.811$\pm$0.041  \\ 
FX Ser &  18040$-$0941  &  2  &  54396.9368  &  0.263$\pm$0.021  &  $-$0.433$\pm$0.022  &  $-$0.969$\pm$0.022  &  $-$1.505$\pm$0.041  \\ 

DR Ser &  18448$+$0523  &  1  &  54039.4356  &  1.344$\pm$0.037  &  1.505$\pm$0.043  &  1.402$\pm$0.040  &  1.051$\pm$0.043  \\ 
DR Ser &  18448$+$0523  &  2  &  54229.1684  &  1.442$\pm$0.041  &  1.572$\pm$0.046  &  1.505$\pm$0.065  &  1.156$\pm$0.063  \\ 

S Sct  &  18476$-$0758  &  1  &  54005.8936  &  $-$0.058$\pm$0.021  &   0.096$\pm$0.024  &  0.044$\pm$0.028  &  $-$0.338$\pm$0.020  \\ 
S Sct  &  18476$-$0758  &  2  &  54396.9396  &  $-$0.099$\pm$0.020  &  0.073$\pm$0.023  &  0.028$\pm$0.028  &  $-$0.385$\pm$0.038  \\ 

RS Cyg &  20115$+$3834  &  1  &  54036.9492  &  0.963$\pm$0.026  &  1.080$\pm$0.029  &  0.867$\pm$0.024  &  0.764$\pm$0.055  \\
RS Cyg &  20115$+$3834  &  2  &  54286.3299  &  1.254$\pm$0.034  &  1.065$\pm$0.043  &  0.742$\pm$0.022  &  0.867$\pm$0.048  \\ 

T Ind  &  21168$-$4514  &  1  &  54069.0284  &  0.242$\pm$0.020  &  0.380$\pm$0.015  &  0.177$\pm$0.019  &  0.005$\pm$0.022  \\ 
T Ind  &  21168$-$4514  &  2  &  54430.9820  &  0.263$\pm$0.021  &  0.387$\pm$0.023  &  0.216$\pm$0.020  &  0.028$\pm$0.017  \\ 

%%
%%\cutinhead{S-RICH STARS}
\hline
\multicolumn{8}{c}{S-RICH STARS} \\
\hline

R Gem  &  07043$+$2246  &  1  &  54068.8159  &  1.210$\pm$0.027  &  1.214$\pm$0.027  &  0.891$\pm$0.037  &  0.476$\pm$0.051  \\
R Gem  &  07043$+$2246  &  2  &  54228.7301  &  1.140$\pm$0.031  &  1.080$\pm$0.029  &  0.786$\pm$0.022  &  0.468$\pm$0.033  \\

NQ Pup &  07507$-$1129  &  1  &  54069.3910  &  1.990$\pm$0.034  &  2.190$\pm$0.041  &  2.024$\pm$0.065  &  1.990$\pm$0.034  \\ 
NQ Pup &  07507$-$1129  &  2  &  54228.8126  &  1.990$\pm$0.034  &  2.215$\pm$0.042  &  2.024$\pm$0.070  &  1.990$\pm$0.068  \\ 

S UMa  &  12417$+$6121  &  1  &  54069.7437  &  2.620$\pm$0.037  &  2.834$\pm$0.030  &  2.627$\pm$0.024  &  2.468$\pm$0.053  \\  
S UMa  &  12417$+$6121  &  2  &  53889.3814  &  3.002$\pm$0.035  &  3.189$\pm$0.041  &  3.002$\pm$0.052  &  2.798$\pm$0.072  \\ 

R Cyg  &  19354$+$5005  &  1  &  54005.9052  &  0.067$\pm$0.017  &  0.022$\pm$0.017  &  $-$0.377$\pm$0.015  &  $-$0.856$\pm$0.049  \\
R Cyg  &  19354$+$5005  &  2  &  54286.3666  &  1.124$\pm$0.031  &  1.009$\pm$0.041  &  0.476$\pm$0.034  &  $-$0.079$\pm$0.051  \\

AA Cyg &  20026$+$3640  &  1  &  54065.7245  &  0.202$\pm$0.013  &  0.528$\pm$0.018  &  0.236$\pm$0.020  &  0.139$\pm$0.019  \\
AA Cyg &  20026$+$3640  &  2  &  54286.3611  &  0.216$\pm$0.020  &  0.573$\pm$0.018  &  0.256$\pm$0.021  &  0.177$\pm$0.032  \\ 

X Aqr  &  22159$-$2109  &  1  &  54286.3244  &  2.190$\pm$0.065  &  2.135$\pm$0.062  &  1.911$\pm$0.044  &  1.721$\pm$0.053  \\ 
X Aqr  &  22159$-$2109  &  2  &  54062.9024  &  2.729$\pm$0.047  &  2.677$\pm$0.045  &  2.447$\pm$0.052  &  2.215$\pm$0.042  \\ 

HR Peg &  22521$+$1640  &  1  &  53928.4751  &  0.649$\pm$0.020  &  0.891$\pm$0.025  &  0.710$\pm$0.031  &  0.620$\pm$0.038  \\ 
HR Peg &  22521$+$1640  &  2  &  54095.8618  &  0.639$\pm$0.020  &  0.843$\pm$0.024  &  0.669$\pm$0.030  &  0.582$\pm$0.028  \\ 

%%
%%\cutinhead{SUPERGIANTS}
\hline
\multicolumn{8}{c}{SUPERGIANTS} \\
\hline
XX Per &  01597$+$5459  &  1  &  54005.9244  &  1.171$\pm$0.032  &  1.088$\pm$0.030  &  0.721$\pm$0.032  &  0.085$\pm$0.059  \\ 
XX Per &  01597$+$5459  &  2  &  54150.6211  &  1.220$\pm$0.033  &  1.156$\pm$0.032  &  0.786$\pm$0.034  &  0.145$\pm$0.062  \\ 

W Per  &  02469$+$5646  &  1  &  54005.9295  &  1.237$\pm$0.036  &  1.171$\pm$0.032  &  0.808$\pm$0.034  &  0.177$\pm$0.064  \\ 
W Per  &  02469$+$5646  &  2  &  54152.9557  &  1.244$\pm$0.034  &  1.178$\pm$0.032  &  0.820$\pm$0.035  &  0.114$\pm$0.060  \\ 

NO Aur &  05374$+$3153  &  1  &  54396.8402  &  0.610$\pm$0.019  &  0.843$\pm$0.024  &  0.582$\pm$0.019  &  0.320$\pm$0.029  \\ 
NO Aur &  05374$+$3153  &  2  &  54188.5900  &  0.620$\pm$0.019  &  0.843$\pm$0.024  &  0.591$\pm$0.019  &  0.312$\pm$0.022  \\

U Lac  &  22456$+$5453  &  1  &  54005.9181  &  0.786$\pm$0.022  &  0.573$\pm$0.018  &  0.196$\pm$0.020  &  $-$0.462$\pm$0.043  \\ 
U Lac  &  22456$+$5453  &  2  &  54328.9353  &  0.764$\pm$0.022  &  0.564$\pm$0.018  &  0.183$\pm$0.019  &  $-$0.440$\pm$0.024  \\ 

NML Cyg$^*$  &  R20445$+$3955  &  1  &  53213.2738  &  
$-$1.53$\pm$0.15 & $-$2.42$\pm$0.15 & ... & $-$4.53$\pm$0.22 \\ 

VX Sgr$^*$  &  18050$-$2213  &  1  &  53634.9152  &  
$-$1.26$\pm$0.15 & $-$1.83$\pm$0.15 & ... & $-$3.36$\pm$0.15 \\ 

\hline
\end{tabular}
\label{phot_table}
\begin{tabular}{l}
$^a$ MDJ = JD$-$2,400,000.5 \\ 
$^*$ only one epoch; photometry from \citet[][see Table 4.4]{sch07} \\
\end{tabular}
\end{table*}

\clearpage

\begin{table*}
\caption{Mira/SR P-L sequence}
\begin{tabular}{lccl}
\hline\hline
Name & Type & Sequence & Notes \\
\hline
\multicolumn{4}{c}{O-RICH STARS} \\
\hline
KU And & M & \emph{D} & low [3.6], large IR excess -- extinction from the circumstellar envelope? \\ 
RW And & M & \emph{C} & along the bottom of the sequence \\ 
VY Cas & SRb & \emph{B}$-$\emph{C'} &  \\ 
SV Psc & SRb & \emph{C'} &  \\ 
RR Per & M & \emph{C} &  \\ 
RV Cam & SRb & \emph{C'} &  \\ 
ET Vir & SRb & \emph{C} & smallest 3.6 \micron\ magnitude  \\ 
RW Boo & SRb & \emph{C} & at the bottom of sequence \emph{C}, small IR excess \\
AX Sco & SRb & \emph{C} & near the top of sequence \emph{C} \\ 
%%
%TV Dra & SR & ... & no period available \\ 
%%
V438 Oph & SRb & \emph{C'}$-$\emph{C} &  \\ 
%%
%TY Dra & Lb & ... & no period available \\ 
%%
%CZ Ser & Lb & ... & no period available \\ 
%%
FI Lyr & SRb & \emph{B} & at the top of sequence \emph{B} \\ 
%%
%V1351 Cyg & Lb & ... & no period available \\ 
%%
Z Cyg & M & \emph{C} & large IR excess \\
V1300 Aql & M & \emph{C}$-$\emph{D} & large IR excess, low 3.6 \micron, extinction from the circumstellar envelope? \\ 
%%
%V584 Aql & Lb & ... & no period available \\ 
%%
RX Vul & M & \emph{C} & large IR excess, lies at the bottom of \emph{C} \\ 
UX Cyg & M & \emph{C} & one of largest IR excesses \\ 
SS Peg & M & \emph{C+} & just below sequence \emph{C}, modest 3.6 \micron\ and IR excess compared to other Miras of similar period \\ 
V563 Cas & M & \emph{C}$-$\emph{D} & just below \emph{C}, among largest IR excesses \\ 
\hline
\multicolumn{4}{c}{C-RICH STARS} \\
\hline
VX And & SRa & \emph{C} & \\ 

HV Cas & M & \emph{C} &  \\ 

R For & M & \emph{C'} & highest Mira 3.6 \micron\ magnitude \\ 

%V623 Cas & Lb & ... & no period available \\ 

UV Aur & M & \emph{C} &  \\ 

SY Per & SR & \emph{C} &  \\ 

TU Tau & SRb & \emph{C'} & largest IR excess of the SRs \\

BN Mon & SRb & \emph{C}$-$\emph{D} & falls just below sequence \emph{C} \\ 

%CR Gem & Lb & ... & no period available \\ 

V614 Mon & SRb & \emph{B} &  \\ 

CGCS 6296 & ... & ... & no period available \\ 

SZ Car & SRb & \emph{C} &  \\ 

V CrB & M & \emph{C} &  \\ 

SX Sco & SR & ... & no period available \\ 

%FX Ser & Lb & ... & no period available \\ 

%DR Ser & Lb & ... & no period available \\ 

S Sct & SRb & \emph{B} &  \\ 

RS Cyg & SRa & \emph{C}$-$\emph{D} & one of two SRs below \emph{C} \\ 

T Ind & SRb & \emph{C} &  \\
\hline
\multicolumn{4}{c}{S-RICH STARS} \\
\hline
R Gem & M & \emph{C}$-$\emph{D} & just below \emph{C}, modest 3.6 \micron\ and IR excess compared to other Miras of similar period \\ 

%NQ Pup & Lb & ... & no period available \\ 

S UMa & M & \emph{C} & at the top of sequence \emph{C}, smallest IR excess of the Miras \\ 

R Cyg & M & \emph{C}$-$\emph{D} & slightly closer to \emph{D} than \emph{C} \\ 

AA Cyg & SRb & \emph{C} & at the top of sequence \emph{C} \\

X Aqr & M & \emph{C} & second smallest Mira IR excess \\ 

HR Peg & SRb & \emph{A} & only source in \emph{A} \\ 
\hline
\multicolumn{4}{c}{SUPERGIANTS} \\
\hline
XX Per & SRc & \emph{C'} & SR with large IR excess \\ 

W Per & SRc & \emph{C'} & right along the lower edge of sequence \emph{C'}; SR with large IR excess \\ 

%NO Aur & Lc & ... & no period available \\ 

%U Lac & SRc & ... & no period available \\ 

NML Cyg & SRc & \emph{C'} & brightest supergiant \\ 

VX Sgr & SRc & \emph{B}$-$\emph{C'} & along the boundary between sequences \emph{B} and \emph{C'}; second brightest supergiant \\ 

\hline
\end{tabular}
\label{tab:PL}
%\tablenotetext{a}{add notes here}
\end{table*}

%%=============================================================================

\end{document}